\documentclass[runningheads,a4paper]{llncs}
\usepackage{times}
\usepackage{url}
\usepackage[usestackEOL]{stackengine}
\usepackage{graphics}
\usepackage{rotating}
\usepackage{xspace}
\usepackage{colortbl}
\usepackage{tabularx}
\usepackage{multirow}
\usepackage{dcolumn}
\usepackage{longtable}
\usepackage[usenames,dvipsnames]{xcolor}
\usepackage{enumitem}
\usepackage{multicol}
\usepackage{lipsum}
\usepackage{subcaption}
\setlist{nolistsep}
\usepackage{booktabs}
\usepackage{amsmath}
\usepackage{amssymb}
\usepackage{algorithm}% http://ctan.org/pkg/algorithms
\usepackage{algpseudocode}% http://ctan.org/pkg/algorithmicx
\usepackage{tikz}
\usetikzlibrary{automata}
\usetikzlibrary[arrows,decorations.pathmorphing,backgrounds,positioning,fit,petri]

\definecolor{darkgreen}{rgb}{0.18,.6,.0}

\tikzset{
  bend angle=30,
  ->,
  shorten >=1pt,
  node distance=2cm and 2cm,
  on grid,
  auto,
  initial where=above,
  initial text=,
  initial distance=0.6cm,
  inner sep=0.5mm,
  smallstate/.style={circle,draw},
  rootstate/.style={rectangle,draw},
  openstate/.style={},
  succstate/.style={font={$\surd$}},
  bendanglelarge/.style={bend angle=45},
  bendanglesmall/.style={bend angle=10},
  dim/.style={lightgray},
  diag/.style={red},
  bisim/.style={red,dashed,-,bend left},
  loop above/.style={in=110,out=70,loop,distance=0.5cm},
  loop left/.style={in=200,out=160,loop,swap,distance=0.5cm},
  loop right/.style={in=20,out=-20,loop,distance=0.5cm},
  loop below/.style={in=290,out=250,loop,distance=0.5cm}
}

\begin{document}

\mainmatter

\title{Taming Asynchrony for \\Attractor Detection in Large Boolean Networks\\(Technical Report)}

\titlerunning{Taming Asynchrony for Attractor Detection in Large Boolean Networks}

%\author{Andrzej Mizera \and Jun Pang \and Qixia Yuan
%\thanks{Supported
%by the National Research Fund, Luxembourg (grant 7814267). }
%}
%\authorrunning{Mizera, Pang and Yuan}
%
%\institute{
%Computer Science and Communications, University of Luxembourg, Luxembourg\\
%\email{firstname.lastname@uni.lu}
%}

%\author{Andrzej Mizera\inst{1} \and Jun Pang\inst{1,2} \and Hongyang Qu\inst{3} \and Qixia Yuan\inst{1}\thanks{Supported
%by the National Research Fund, Luxembourg (grant 7814267). }}
\author{Andrzej Mizera\inst{1} \and Jun Pang\inst{1,2} \and Hongyang Qu\inst{3} \and Qixia Yuan\inst{1} }
\authorrunning{Mizera et al.}
\institute{
Faculty of Science, Technology and Communication , University of Luxembourg
\and
Interdisciplinary Centre for Security, Reliability and Trust, University of Luxembourg\\
\and
Department of Automatic Control \& Systems Engineering, University of Sheffield
%\email{firstname.lastname@uni.lu}
}

\maketitle

\begin{abstract}
Boolean networks is a well-established formalism for modelling biological systems.
A vital challenge for analysing a Boolean network is to identify all the attractors.
This becomes more challenging for large asynchronous Boolean networks,
due to the asynchronous updating scheme.
Existing methods are prohibited due to the well-known state-space explosion problem in large Boolean networks.
In this paper, we tackle this challenge by proposing a SCC-based decomposition method. 
We prove the correctness of our proposed method
and demonstrate its efficiency with two real-life biological networks. 
%
%\keywords{Probabilistic Boolean networks, Markov chains, steady-state analysis, approximation}
\end{abstract}

%===============================================
\section{Introduction}
\label{sec:intro}
%===============================================

Boolean networks (BNs) is a~well-established framework used for modelling biological systems such
as gene regulatory networks (GRNs). It has the advantage of being simple yet able to capture the
important dynamic properties of the modelled system, e.g., the system's \emph{attractors}.
An~attractor of a~biological system is a~set of the system's states satisfying that any two states in this set
can be reached from each other and the system remains in this set until some external stimulus
pushes the system out of it. Attractors are hypothesised to characterise cellular
phenotypes~\cite{KS69} or to correspond to functional cellular states such as proliferation,
apoptosis, or differentiation~\cite{HS01}. For example, in the study of Sanchez-Corrales et
al.~\cite{SAM10}, attractors of an~\emph{Arabidopsis thaliana} system correspond to stable gene
expression levels during the different stages of flower development. Identification of attractors
is therefore of great importance for the analysis of biological systems modelled as BNs.

%A BN is simple in the sense that it usually contains enumerable number of nodes;
%however,
Attractor detection of a BN is non-trivial since attractors
are determined based on the BN's states, the number of which is exponential in the number of nodes.
%Under the BN framework,
A~lot of efforts have been put in the development of attractor detection
algorithms and tools. In the early 2000s, an~enumeration and simulation method has been proposed.
The idea is to enumerate all the possible states and to run simulation from each of them until
an~attractor is found~\cite{SG01}.
This method is largely restricted by the network size since the time grows exponentially with the number of nodes.
In 2006, Irons proposed
a~method to detect BNs with a special topology~\cite{irons06}, making it possible to deal with BNs with maximum 50 nodes.
Later on, the performance has been greatly improved with
two techniques, i.e., binary decision diagrams (BDDs) and satisfiability (SAT) solvers.
%BDD is a~compact data structure for representing a~Boolean function.
BDD-based methods~\cite{ALG07,AAG08} encode Boolean functions of BNs with BDDs, use BDD operations
to capture the dynamics of the network, and use BDD structure to represent the network's
corresponding transition system.
Using the BDD operations, the forward and backward reachable states can be often efficiently computed.
Detecting attractors is then reduced to finding fix point set of states in the corresponding
transition system. The other technique transforms attractor detection in BNs into a~SAT
problem~\cite{DT11}. An~unfolding of the transition relation of the BN for a~bounded number of
steps is represented as a~propositional formula. The formula is then solved by a~SAT solver to
identify a~valid path in the state transition system of the BN. The process is repeated iteratively
for larger bounded numbers of steps until all attractors are identified.
%The efficiency of the algorithm largely relies on the number of unfolding steps required and the number of nodes in the BN.
Recently, a~few decomposition methods~\cite{ZKF13,GYW14,YQP16} were proposed
to deal with large BNs. The main idea is to decompose a~large BN into small components based on
its structure, detect attractors in the small components, and then recover the attractors of the original BN.
%Since the decomposition is performed in the BN structure,
%the decomposition time cost is linear to the number of nodes.

\vspace{-0.2mm}
The above mentioned methods are mainly designed for BNs with the \emph{synchronous} updating
scheme, i.e., BNs where the values of all the nodes are updated simultaneously. In biology,
however, the update speed of each node is not necessarily the same. Updating nodes values
\emph{asynchronously} is
considered more realistic. In synchronous BNs, an~attractor is either a~single state selfloop
or a~cycle since there is exactly one outgoing transition for each state. Under the
asynchronous updating scheme, each state may have multiple outgoing transitions. Therefore,
an~attractor in general is a~bottom strongly connected component (BSCC)
%\footnote{It is also referred as \emph{loose attractor} in the literature~\cite{WSA12}.}
in the corresponding state transition system.
The potentially complex attractor structure renders SAT-based methods
ineffective as the respective SAT formulas become prohibitively large. Besides, the decomposition
methods~\cite{ZKF13,GYW14,YQP16} are also prohibited by the asynchronous updating requirement.
Moreover, BDD-based methods face the state-space explosion problem even in the synchronous updating
scheme. In the asynchronous updating scheme, the problem gets even worse
%largest model the BDD-based methods can handle is even smaller due to the fact
as the number of edges in the state transition system increases multiple times.

\vspace{-0.2mm}
In this paper, we tackle the challenge of attractor detection for asynchronous BNs, especially for
large ones, and we propose a~strongly connected component (SCC) based decomposition method:
%Our idea is inspired by the decomposition methods for synchronous BNs:
decompose a~BN into sub-networks called \emph{blocks} according to the SCCs in the BN and recover attractors of the original BN based on attractors of the blocks.
Since the \emph{decomposition is performed on the BN structure, not in the state space},
the decomposition time cost is linear in the number of nodes and the state space of each block is exponentially smaller in comparison to that of the original BN.
Our method shares similar ideas on the way of decomposition as those used for synchronous BNs.
However, due to the asynchronous updating scheme, the \emph{bottom-up} decomposition methods~\cite{ZKF13,GYW14}
for synchronous BNs are no longer valid, as they
may produce spurious attractors.
%\footnote{For the sake of brevity, detailed explanations are omitted from the current paper.}
The asynchrony poses two main challenges for the decomposition methods: one is to
take care of the dependency relations between different blocks; the other is to strictly comply with
the asynchronous updating scheme when recovering attractors from different blocks. To overcome
these difficulties, we order the blocks according to their dependency relations and detect
attractors of each block with consideration of the block that it depends on.
In this way, our method is \emph{top-down}, starting with elementary blocks which do not depend on others.
We prove that our proposed method can correctly detect all the attractors of a~BN (Section~\ref{sec:decomposition}),
and we implement it using efficient BDD techniques (Section~\ref{sec:algorithms}). Evaluation results show that our method can
effectively detect attractors of two~real-life biological networks (Section~\ref{sec:eva}).

%===============================================
\section{Preliminaries}
\label{sec:pre}
%===============================================

%===============================================
\subsection{Boolean networks}
\label{ssec:bn}
%===============================================
A BN describes elements of a~biological system with binary-valued nodes and
interactions between elements with Boolean functions. It was first introduced by Kauffman
in 1969 as a~class of simple models for analysing the dynamical properties of GRNs~\cite{Kauffman69a},
in which each gene was assumed to be in only two possible states: ON/OFF.
\begin{definition}[Boolean network]
A~\emph{Boolean network} $G(V, \boldsymbol{f})$ consists of a~set of nodes
$V=\{v_1,v_2,\ldots,v_n\}$, also referred to as genes, and a~vector of Boolean functions
$\boldsymbol{f}=(f_1,f_2,\ldots,f_n)$, where $f_i$ is a~predictor function associated with node
$v_i$ ($i=1,2,\ldots,n$). A~state of the network is given by a~vector
$\boldsymbol{x}=(x_1,x_2,\ldots,x_n) \in \{0,1\}^n$, where $x_i \in \{0,1\}$ is a~value assigned
to node $v_i$.
\end{definition}

Each node $v_i \in V$ has an~associated subset of nodes $\{v_{i_1},v_{i_2},\ldots,v_{i_{k(i)}}\}$,
referred to as the set of \emph{parent nodes} of $v_i$, where $k(i)$ is the number of parent nodes
and $1\leq i_1 < i_2 < \cdots < i_{k(i)} \leq n$. Starting from an~initial state, the BN evolves
in time by transiting from one state to another. The state of the network at a~discrete time point
$t$ ($t=0,1,2,\ldots$) is given by a~vector $\boldsymbol{x}(t)=(x_1(t),x_2(t),\ldots,x_n(t))$,
where $x_i(t)$ is a~binary-valued variable that determines the value of node $v_i$ at time point
$t$. The value of node $v_i$ at time point $t+1$ is given by the predictor function $f_i$ applied
to the values of the parent nodes of $v_i$ at time $t$, i.e., $x_i(t+1) =
f_i(x_{i_1}(t),x_{i_2}(t),\ldots,x_{i_{k(i)}}(t))$. For simplicity, with slight abuse of notation,
we use $f_i(x_{i_1},x_{i_2},\ldots,x_{i_{k(i)}})$ to denote the value of node $v_i$ at the next
time step. For any $j \in [1,k(i)]$, node $v_{i_j}$ is called  a~\emph{parent node} of $v_i$ and
$v_i$ is called a~\emph{child node} of $v_{i_j}$.

In general, the Boolean predictor functions can be formed by combinations of any logical
operators, e.g., logical \textsc{and} $\wedge$, \textsc{or} $\vee$, and \textsc{negation} $\neg$,
applied to variables associated with the respective parent nodes. The BNs are divided into two
types based on the time evolution of their states, i.e., {\em synchronous} and {\em asynchronous}.
In synchronous BNs, values of all the variables are updated simultaneously; while in asynchronous
BNs, one variable is updated at a~time. The synchronous updating scheme is used mostly due to its
simplicity; however, for the modelling of GRNs, the asynchronous scheme is
more suitable as the expression of a~gene is usually not an~instantaneous process.

In this paper, we focus on asynchronous BNs. The transition relation of an~asynchronous BN is
given by $T\left(\boldsymbol{x}(t),\boldsymbol{x}(t+1)\right)=$
\begin{multline}
\label{equ:transition}
\left(x_i(t+1)\leftrightarrow f_i(x_{i_1}(t),x_{i_2}(t),\cdots,x_{i_{k_i}}(t))\right)
\bigwedge_{j=1,j\neq i}^{n} (x_j(t+1)\leftrightarrow x_j(t)) .
\end{multline}
It states that node $v_i$ is updated by its Boolean function and other nodes are kept unchanged.
Each node has a~chance to be updated by its Boolean function, therefore there are $n$ outgoing
transitions in maximum from any state.

Many key characters of a~BN, e.g., attractors, are often examined in the level of its state transition system (STS).
\footnote{For presentation purpose, a few state transition systems are drawn as graphs in the remaining part of the paper
and are also referred to as transition graphs.}
Formally, the state transition system and attractors of a BN are defined as follows.

\begin{definition}[State transition system]
A~state transition system $\mathcal{T}$ is a~3-tuple $\langle
S,S_0,\allowbreak T\rangle$ where $S$ is a~finite set of states, $S_0
\subseteq S$ is the initial set of states and $T \subseteq S \times S$
is the transition relation.
%specifying the evolvement of the system.
When $S=S_0$, we write $\langle S, \allowbreak T\rangle$.
\end{definition}

An asynchronous BN can be easily modelled as a~state transition system:
the set $S$ is just the state space of the BN so there are $2^n$ states for a~BN with $n$ nodes;
the initial states set $S_0$ is the same as $S$ since usually all states are accessible in a biological system modelled as a BN;
finally, the transition relation $T$ is given by Equation~\ref{equ:transition}.

\begin{definition}[Attractor of a~BN]
An~\emph{attractor} of a~BN is a~set of states satisfying that any state in this set can be
reached from any other state in this set and no state in this set can reach any other state that
is not in this set.
\end{definition}

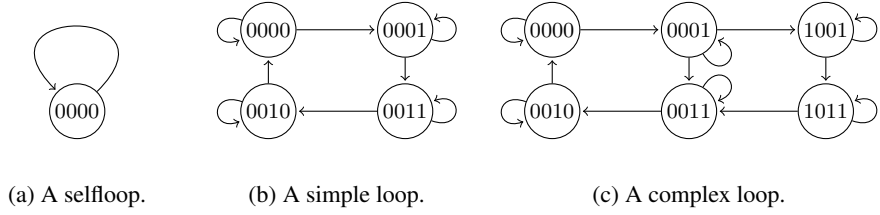
\begin{figure}[t!]
\begin{subfigure}[b]{0.25\textwidth}
\begin{center}
\begin{tikzpicture}[scale=0.9, every node/.style={transform shape}]
\node[smallstate]  (r) {$0000$};

 \path     (r)  edge [loop] node {} (r);
\end{tikzpicture}
\end{center}
\caption{A selfloop.}
\label{fig:selfloop}
    \end{subfigure}
    \begin{subfigure}[b]{0.3\textwidth}
\begin{center}
\begin{tikzpicture}[scale=0.9, every node/.style={transform shape}]
 \node[smallstate]  (r1) {$0000$};
 \node[smallstate] (r2)   [right=of r1]       {$0001$};
 \node[smallstate]  (r3)  at(2,-1.2)   {$0011$};
 \node[smallstate]  (r4)  at(0,-1.2)   {$0010$};

 \path     (r1)  edge [] node {} (r2)
 			  (r1)  edge [loop left] node {} (r1)
			  (r2)  edge [] node {} (r3)
			  (r2)  edge [loop right] node {} (r2)
			  (r3)  edge [] node {} (r4)
			  (r3)  edge [loop right] node {} (r3)
			  (r4)  edge [] node {} (r1)
			  (r4)  edge [loop left] node {} (r4);
\end{tikzpicture}
\end{center}
\caption{A simple loop.}
\label{fig:simpleloop}
\end{subfigure}
    \begin{subfigure}[b]{0.45\textwidth}
\begin{center}
\begin{tikzpicture}[scale=0.9, every node/.style={transform shape}]
 \node[smallstate]  (r1) {$0000$};
 \node[smallstate]  (r2) [right =of r1] {$0001$};
 \node[smallstate]  (r3) at(2,-1.2) {$0011$};
 \node[smallstate]  (r4) at(0,-1.2) {$0010$};
 \node[smallstate]  (r5) [right =of r2] {$1001$};
 \node[smallstate]  (r6) at(4,-1.2) {$1011$};

 \path     (r1)  edge [] node {} (r2)
 			  (r1)  edge [loop left] node {} (r1)
			  (r2)  edge [] node {} (r3)
			  (r2)  edge [in=-20,out=-60,loop,distance=0.5cm] node {} (r2)
			  (r3)  edge [] node {} (r4)
			  (r3)  edge [in=20,out=60,loop,distance=0.5cm] node {} (r3)
			  (r4)  edge [] node {} (r1)
			  (r4)  edge [loop left] node {} (r4)
			  (r2)  edge [] node {} (r5)
			  (r5)  edge [loop right] node {} (r5)
			  (r5)  edge [] node {} (r6)
			  (r6)  edge [] node {} (r3)
			  (r6)  edge [loop right] node {} (r6)
			  ;
\end{tikzpicture}
\end{center}
\caption{A complex loop.}
\label{fig:complexloop}
\end{subfigure}
\caption{Three types of attractors in an asynchronous BN.}
\label{fig:bn_attr}
\vspace{-3mm}
\end{figure}

The attractors of a~BN characterise its long-run behaviour~\cite{SD10} and are of particular interest
due to their biological interpretation
as, for instance, attractors are hypothesised to characterise cellular phenotypes~\cite{KS69}.
When analysing an~attractor, we often need to identify transition relations between the attractor
states. We refer to an~attractor together with its state transition relations as
an~\emph{attractor system}.
%Depending on the number of states, attractors can be divided into two types: \emph{singleton
%attractors} containing only one state; \emph{complex attractors} having more than one state.
The states constituting an~attractor are called \emph{attractor states}.
In the synchronous updating scheme, each state can only have one outgoing transition.
%Therefore, the attractor system of an~attractor in a~synchronous BN is simply a~loop.
Therefore, the attractor system in a~synchronous BN is simply a~loop.
By detecting all the loops in a~synchronous BN, one can identify all its attractors.
However, it becomes much more complicated with the asynchronous updating scheme.
%The attractor system of an~attractor does not necessarily need to be a~loop and may have a~more intricate topology.
The attractor system does not necessarily need to be a~loop and may have a~more intricate topology.
In fact, it may include several loops. We list three general types of attractors of an~asynchronous BN:
a~singleton attractor, i.e., a~\emph{selfloop}, as shown in Figure~\ref{fig:selfloop};
a~\emph{simple loop}, as shown in Figure~\ref{fig:simpleloop}; and a~\emph{complex loop}, as shown
in Figure~\ref{fig:complexloop}. The selfloops and simple loops also exist in the corresponding
synchronous BN. %(note that simple loop attractors do not contain selfloop edges in synchronous BNs).
Therefore, one can identify the selfloops and simple loop attractors of
an~asynchronous BN by detecting the attractors of its corresponding synchronous BN.\footnote{
Note that some of the detected attractors in the form of loops in a synchronous BN
can be absent in the corresponding asynchronous BN. See~\cite{WSA12,AAG08} for detailed discussions.}
However, this is not the case for complex loop attractors. Special algorithms need to be designed to detect such
attractors under asynchronous updating scheme.

%===============================================
\subsection{Encoding BNs in BDDs}
\label{ssec:encoding}
%===============================================
Binary decision diagrams (BDDs) were introduced
%by Lee in~\cite{LCY59} and Akers in~\cite{ASB78}
to represent Boolean functions~\cite{LCY59,ASB78}. BDDs have the advantage of memory efficiency and
have been applied in model checking algorithms to alleviate the state space explosion problem.
A~BN $G(V, \boldsymbol{f})$ can be easily encoded in a~BDD by modelling a BN as an STS.
Each variable in $V$ can be represented by a~binary BDD variable. By
slight abuse of notation, we use $V$ to denote the set of BDD
variables. In order to encode the transition relation, another set
$V'$ of BDD variables, which is a~copy of $V$, is introduced: $V$
encodes the possible current states, i.e., $\boldsymbol{x}(t)$, and
$V'$ encodes the possible next states, i.e.,
$\boldsymbol{x}(t+1)$. Hence, the transition relation can be viewed as
a~Boolean function $T: 2^{|V|+|V'|} \to \{0,1\}$, where values $1$ and
$0$ indicate a~valid and an~invalid transition, respectively. Our
attractor detection algorithms also use two basis functions: ${\it
  Image}(X, T)=\{s'\in S\mid \exists s\in X \mbox{ such that } (s, s')
\in T\}$, which returns the set of target states that can be reached
from any state in $X\subseteq S$ with a~single transition in $T$;
${\it Preimage}(X, T)=\{s'\in S\mid \exists s\in X \mbox{ such that }
(s', s) \in T\}$, which returns the set of predecessor states that can
reach a state in $X$ with a single transition. To simplify the
presentation, we also define ${\it Preimage}^i(X, T)=\underbrace{{\it
Preimage}(...({\it Preimage}(X, T)))}_{i}$ with ${\it
Preimage}^0(X, T)=X$.  Thus, the set of all states that can reach a
state in $X$ via transitions in $T$ is defined as a fix point ${\it
Predecessors}(X, T)=\bigcup\limits_{i=0}^n{\it Preimage}^n(X, T)$
such that ${\it Preimage}^n(X, T)= {\it Preimage}^{n+1}(X, T)$. Given
a set of states $X\subseteq S$, the projection $T|_X$ of $T$ on $X$ is
defined as $T|_X=\{(s, s')\in T \mid s\in X \land s'\in X\}$.

%===============================================
\section{Method}
\label{sec:decomposition}
%===============================================
%As the number of states of a BN is exponential to the number of nodes in this BN.
%Algorithm~\ref{alg:detect} face the state space explosion problem when the BN becomes large.
%We propose a decomposition method for dealing with large BNs.

In this section, we describe in details our SCC-based decomposition method for detecting
attractors of large asynchronous BNs and prove its correctness. The method consists of three main
steps. First, we divide a~BN into sub-networks called \emph{blocks}.
This step is performed based on the BN network structure
which contains enumerable number of nodes
and therefore it can be executed efficiently.
Second, we detect attractors of each block. This step is performed on the constructed STSs of the
blocks. Notice that for each block the size of its STS is exponentially reduced with respect to
the size of the STS of the original BN. Finally, we recover attractors of the original BN by
merging the detected attractors of the blocks.
%
%===============================================
\subsection{Decomposition a BN into blocks}
\label{ssec:decomposition}
%===============================================
%The decomposition is performed in the network structure which contains enumerable nodes.
%Therefore, the time cost for decomposition is negligible.
We start a~detailed presentation of our approach by giving the formal definition of a~block.

\begin{definition}[Block]
Given a~BN $G(V,\boldsymbol{f})$ with $V=\{v_1,v_2,\ldots,v_n\}$ and
$\boldsymbol{f}=\{f_1,\allowbreak f_2,\ldots,f_n\}$, a~\emph{block} $B(V^{B},\boldsymbol{f^{B}})$
is a~subset of the network, where $V^{B} \subseteq V$. For any node $v_i \in V^{B}$, if $B$
contains all the parent nodes of $v_i$, its Boolean function in $B$ remains the same as in $G$,
i.e., $f_i$; otherwise, the Boolean function is undetermined, meaning that additional information
is required to determine the value of $v_i$ in $B$. We call the nodes with undetermined Boolean
functions as \emph{undetermined nodes}.
%Note that the undetermined Boolean function does not change the asynchronous updating mode.
We refer to a~block as an~\emph{elementary block} if it contains no undetermined nodes.
\end{definition}

We consider asynchronous networks in this paper and therefore a~block is also under the
asynchronous updating scheme, i.e., only one node in the block can be updated at any given time
point no matter this node is undetermined or not.

%\begin{example}
%\label{ex:block}
%Consider the BN given in Example~\ref{ex:bn}. Nodes $v_1$ and $v_2$ together with their Boolean
%functions form a~block of this BN. In this block, all the nodes are determined nodes. Therefore,
%it is an~elementary block.
%\end{example}

We now introduce a~method to construct blocks using SCC-based decomposition. Formally, the
standard graph-theoretical definition of an~SCC is as follows.
\begin{definition}[SCC]
Let $\mathcal{G}$ be a~directed graph and $\mathcal{V}$ be its vertices. A~strongly connected
component (SCC) of $\mathcal{G}$ is a~maximal set of vertices $C \subseteq \mathcal{V}$ such that
for every pair of vertices $u$ and $v$ in $C$, there is a~directed path from $u$ to $v$ and vice
versa.
\end{definition}

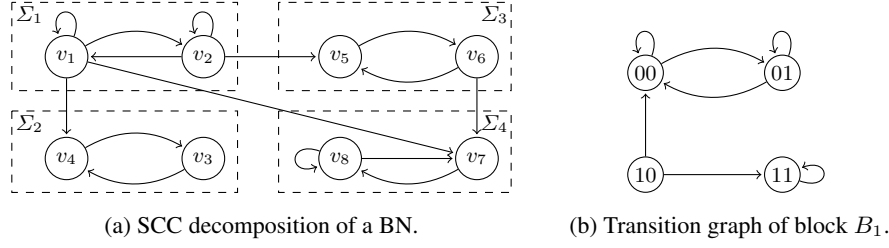
\begin{figure}[t!]
 \begin{subfigure}[b]{0.6\textwidth}
\begin{center}
\begin{tikzpicture}[scale=0.9, every node/.style={transform shape}]
 \node[smallstate] (s1)                       {$~v_1~$};
 \node[smallstate]         (s2)  [right=of s1]   {$~v_2~$};
 \node[smallstate]         (s3)  at(2,-1.5) {$~v_3~$};
 \node[smallstate]         (s4)  at(0,-1.5)  {$~v_4~$};
  \node[smallstate]         (s5)  [right=of s2]   {$~v_5~$};
   \node[smallstate]         (s6)  [right=of s5]   {$~v_6~$};
 \node[smallstate]         (s7)  at(6,-1.5)   {$~v_7~$};
 \node[smallstate]         (s8)  [left=of s7]  {$~v_8~$};
 \draw[dashed] (-0.8,0.8) rectangle (2.5,-0.5);
 \draw[dashed] (3.1,0.8) rectangle (6.5,-0.5);
 \draw[dashed] (-0.8,-0.8) rectangle (2.5,-2);
 \draw[dashed] (3.1,-0.8) rectangle (6.5,-2);
 \node[anchor=north west] at (-0.8,0.8) {$\Sigma_1$};
 \node[anchor=north west] at (6,0.8) {$\Sigma_3$};
 \node[anchor=north west] at (-0.8,-0.8) {$\Sigma_2$};
 \node[anchor=north west] at (6,-0.8) {$\Sigma_4$};
 \path     (s1)  edge [loop above] node {} (s1)
 (s1)  edge [bend left] node {} (s2)
  (s1)  edge [] node {} (s4)
  (s1)  edge [] node {} (s7)
   (s2)  edge [loop above] node {} (s2)
  (s2)  edge [] node {} (s1)
   (s2)  edge [] node {} (s5)
   (s3)  edge [bend left] node {} (s4)
     (s4)  edge [bend left] node {} (s3)
   (s5)  edge [bend left] node {} (s6)
   (s6)  edge [bend left] node {} (s5)
   (s6)  edge [] node {} (s7)
   (s7)  edge [bend left] node {} (s8)
   (s8)  edge [] node {} (s7)
   (s8)  edge [loop left] node {} (s8) ;
\end{tikzpicture}
\end{center}
\vspace{-3mm}
\caption{SCC decomposition of a BN.}
\label{fig:sccde}
\end{subfigure}
\begin{subfigure}[b]{0.4\textwidth}
 \begin{center}
 \begin{tikzpicture}[scale=0.9, every node/.style={transform shape}]
 \node[smallstate] (s1)                       {$00$};
 \node[smallstate]         (s2)  [right=of s1]   {$01$};
 \node[smallstate]         (s3)  at(0,-1.5)  {$10$};
 \node[smallstate]         (s4)  [right =of s3]  {$11$};

 \path     (s1)  edge [loop above] node {} (s1)
 		   (s1)  edge [bend left] node {} (s2)
 		    (s2)  edge [loop above] node {} (s2)
 		     (s2)  edge [bend left] node {} (s1)
           (s3) edge [] node {} (s1)
           (s3)  edge [] node {} (s4)
           (s4)  edge [loop right]       node {} (s4)
           ;
\end{tikzpicture}
\end{center}
\vspace{-3mm}
\caption{Transition graph of block $B_1$.}
\label{fig:tgb}
\end{subfigure}
\caption{An example of the SCC decomposition and the transition graph of block $B_1$.}
\label{fig:sccandtg}
\vspace{-3mm}
\end{figure}

We first decompose a~given BN, its network structure, into SCCs. Figure~\ref{fig:sccde} shows the
decomposition of a~BN into four SCCs: $\Sigma_1$, $\Sigma_2$,  $\Sigma_3$, and $\Sigma_4$. A~node
outside an~SCC that is a~parent to a~node in the SCC is referred to as a~\textit{control node} of
this SCC. In Figure~\ref{fig:sccde}, node $v_1$ is a~control node of $\Sigma_2$ and $\Sigma_4$;
node $v_2$ is a~control node of $\Sigma_3$; and node $v_6$ is a~control node of $\Sigma_4$. The
SCC $\Sigma_1$ does not have any control node.

\begin{definition}[Parent SCC, Ancestor SCC]
An~SCC $\Sigma_i$ is called a~\emph{parent SCC} (or \emph{parent} for short) of another SCC
$\Sigma_j$ if $\Sigma_i$ contains at least one control node of $\Sigma_j$. Denote $P(\Sigma_i)$
the set of parent SCCs of $\Sigma_i$. An~SCC $\Sigma_k$ is called an~\emph{ancestor SCC} (or
\emph{ancestor} for short) of an~SCC $\Sigma_j$ if and only if either (1) $\Sigma_k$ is a~parent
of $\Sigma_j$ or (2) $\Sigma_k$ is a~parent of $\Sigma_j$'s ancestor. Denote $\Omega(\Sigma_j)$
the set of ancestor SCCs of $\Sigma_i$.
\end{definition}

An~SCC together with its control nodes forms a~\textit{block}. For example, in
Figure~\ref{fig:sccde}, $\Sigma_2$ and its control node $v_1$ form one block $B_2$. $\Sigma_1$
itself is a~block, denoted as $B_1$, since the SCC it contains does not have any control node. If
a~control node in a~block $B_i$ is a~determined node in another block $B_j$, block $B_j$ is called
a~\emph{parent} of block $B_i$ and $B_i$ is a~child of $B_j$. The concepts of parent and ancestor
are naturally extended to blocks.

By adding directed edges from all parent blocks to all their child blocks, we form a~directed
acyclic graph (DAG) of the blocks as the blocks are formed from SCCs. We notice here that in our
decomposition approach, as long as the block graph is guaranteed to be a~DAG, other strategies to
form blocks can be used.

Two blocks can be merged into one larger block. For example, blocks $B_1$ and $B_2$ can be merged
together to form a~larger block $B_{1,2}$.

A~state of a~block is a~binary vector of length equal to the size of the block which determines
the values of all the nodes in the block.
%Let us denote by $X$ and $X^B$ the state spaces of a~BN
%$G$ and one of its blocks, $B$, respectively.
In this paper, we use a~number of operations on the states of a~BN and its blocks. Their
definitions are given below.

\begin{definition}[Projection map, Compressed state, Mirror states]
\label{def:mirror}
For a~BN $G$ and its block $B$, where the set of nodes in $B$ is $V^B=\{v_1,v_2,\ldots,v_m\}$
and the set of nodes in $G$ is $V=\{v_1,v_2,\ldots,v_m,v_{m+1},\ldots, v_n\}$, the
\emph{projection map} $\pi_{B}: X \to X^{B}$ is given by
$\boldsymbol{x}=\left(x_1,x_2,\ldots,x_m,x_{m+1},\ldots,x_n\right) \mapsto
\pi_{B}(\boldsymbol{x})=(x_1,x_2,\ldots,x_m)$.
For any set of states $S \subseteq X$, we define $\pi_B(S)=\{\pi_B(\boldsymbol{x}):\boldsymbol{x}\in
S\}$. The projected state $\pi_B(\boldsymbol{x})$ is called a~\emph{compressed state} of
$\boldsymbol{x}$. For any state $\boldsymbol{x}^B \in X^B$, we define its set of \emph{mirror
states} in $G$ as $\mathcal{M}_{G}(\boldsymbol{x}^B)=\{\boldsymbol{x} ~|~ \pi_{B}(\boldsymbol{x})
= \boldsymbol{x}^B\}$. For any set of states $S^B\subseteq X^B$, its set of mirror states is
$\mathcal{M}_{G}(S^B)=\{\boldsymbol{x}~|~\pi_B(\boldsymbol{x})\in S^B\}$.
\end{definition}

The concept of the projection map can be extended to blocks. Given a~block
with nodes $V^B=\{v_1,v_2,\ldots,v_m\}$, let
$V^{B'}=\{v_1,v_2,\ldots,v_j\}\subseteq V^B$. We can define $\pi_{B'}: X^B \to X^{B'}$ as
$\boldsymbol{x}=\left(x_1,x_2,\ldots,x_m\right) \mapsto
\pi_{B'}(\boldsymbol{x})=(x_1,x_2,\ldots,x_j)$ and for a~set of states $S^B\subseteq X^B$, we
define $\pi_{B'}(S^B)=\{\pi_{B'}(\boldsymbol{x}):\boldsymbol{x}\in S^B\}$.

%===============================================
\subsection{Detection of attractors in blocks}
\label{ssec:detectionblock}
%===============================================

An~elementary block does not depend on any other block while a~non-elementary block does.
Therefore, they should be treated separately. We first consider the case of elementary blocks.
An~elementary block is in fact a~BN; therefore, the notion of attractors of an~elementary block is
given by the definition of attractors of a~BN.

Next, we introduce the following concept.

\begin{definition}[Preservation of attractors]
\label{def:preserve}
Given a~BN $G$ and an~elementary block $B$ in $G$, let $\mathcal{A}=\{A_1,A_2,\ldots,A_m\}$ be the
set of attractors of $G$ and $\mathcal{A}^{B}=\{A_1^{B},A_2^{B},\ldots,\allowbreak A_{m'}^{B}\}$
be the set of attractors of $B$. We say that $B$ \emph{preserves the attractors} of $G$ if for any
$k \in [1,m]$, there is an~attractor $A^B_{k'} \in \mathcal{A}^B$ such that $\pi_B(A_k) \subseteq
A^B_{k'}$.
\end{definition}

\begin{example}
\label{ex:preserve}
Consider the Boolean network shown in Figure~\ref{fig:sccde}. The Boolean functions of this
network are given as follows:
\begin{equation}
\begin{cases}
  \begin{tabular}{p{2.2cm}p{2.4cm}p{3.1cm}p{2.5cm}}
  $f_1=x_1 \wedge x_2$,  & $f_2=x_1 \vee \neg x_2$, &$ f_3=\neg x_4$,
&$ f_4=x_1 \wedge \neg x_3$,  \\ $f_5=x_2 \wedge x_6$,  &$f_6=x_5$,
 &$f_7=(x_1 \vee x_6) \wedge x_8$,& $f_8=x_7 \vee  x_8.$
  \end{tabular}
\end{cases}
\end{equation}
It has 10 attractors, i.e.,
$\mathcal{A}=
\{\{(0*100000)\},\{(0*100001)\},\{(110100\allowbreak00)\},
\{(1101001\allowbreak1\allowbreak)\allowbreak\},\{(110\allowbreak11100)\},\{(11011111)\},\{(11100000)\},\{(11100011)\},
\{(1\allowbreak1101100)\},\{(11\allowbreak1\allowbreak01111)\}\}$
($*$ means either 0 or 1).
Nodes $v_1$ and $v_2$ form an~elementary
block $B_1$. Since $B_1$ is an~elementary block, it can be viewed as a~BN. The transition graph of
this block is shown in Figure~\ref{fig:tgb}. Its set of attractors is
$\mathcal{A}^{B_1}=\{\{(0*)\},\{(11)\}\}$ (nodes are arranged as $v_1$, $v_2$).
We have $\pi_{B_1}(\{(0*100000)\})=\{(0*)\}\in \mathcal{A}^{B_1}$
and $\pi_{B_1}(\{(0*100001)\})=\{(0*)\}\in \mathcal{A}^{B_1}$.
For the remaining attractors, their compressed set of state is always $\{(11)\}$,
which belongs to $\mathcal{A}$.
Hence, block $B_1$ preserves the attractors of the original BN $G$.
\end{example}

\begin{figure}[t!]
\begin{center}
%\begin{subfigure}[b]{0.35\textwidth}
%\begin{center}
%\begin{tikzpicture}[scale=0.9, every node/.style={transform shape}]
% \node[smallstate] (s1)                       {$~00~$};
% \node[smallstate]         (s2)  [right=of s1]   {$~01~$};
% \node[smallstate]         (s3)  at(2,-1.2)  {$~11~$};
% \node[smallstate]         (s4)  at(0,-1.2)  {$~10~$};
%
% \path     (s1)  edge [loop above] node {} (s1)
% (s1)  edge [bend left] node {} (s2)
%  (s2)  edge [bend left] node {} (s1)
%  (s2)  edge [loop above] node {} (s2)
%   (s3)  edge [] node {} (s2)
%   (s3)  edge [loop right] node {} (s3)
%      (s3)  edge [] node {} (s4)
%   (s4)  edge [loop above] node {} (s4)
%           ;
%\end{tikzpicture}
%\end{center}
%\caption{Transition graph in Example~\ref{ex:preserve}.}
%\label{fig:blockgraph}
%    \end{subfigure}
    \begin{subfigure}[b]{0.6\textwidth}
    \begin{center}
\begin{tikzpicture}[scale=0.8, every node/.style={transform shape}]
 \node[smallstate] (s1)                       {$000$};
 \node[smallstate]         (s2)  [right=of s1]   {$001$};
 \node[smallstate]         (s3)  at(0,-1.1) {$100$};
 \node[smallstate]         (s4)  at(2,-1.1) {$101$};
 \node[smallstate] (s5)         [right=of s2]              {$011$};
 \node[smallstate]         (s6)  [right=of s5]   {$010$};
 \node[smallstate]         (s7)  at(4,-1.1)  {$111$};
 \node[smallstate]         (s8)  at(6,-1.1)  {$110$};

 \path     (s1)  edge [loop left] node {} (s1)
  (s1)  edge [bend left] node {} (s3)
  (s2)  edge [] node {} (s1)
   (s2)  edge [loop above] node {} (s2)
   (s2)  edge [bend left] node {} (s4)
   (s3)  edge [loop left] node {} (s3)
   (s3)  edge [bend left] node {} (s1)
   (s4)  edge [bend left] node {} (s2)
   (s4)  edge [] node {} (s3)
   (s4)  edge [] node {} (s7)
   (s4)  edge [loop below] node {} (s4)
   (s5)  edge [] node {} (s2)
   (s5)  edge [bend left] node {} (s7)
   (s5)  edge [loop above] node {} (s5)
   (s6)  edge [] node {} (s5)
   (s6)  edge [loop right] node {} (s6)
   (s6)  edge [bend right] node {} (s1)
   (s6)  edge [bend left] node {} (s8)
   (s7)  edge [bend left] node {} (s5)
   (s7)  edge [loop below] node {} (s7)
   (s8)  edge [loop right] node {} (s8)
   (s8)  edge [bend left] node {} (s3)
   (s8)  edge [bend left] node {} (s6)
   (s8)  edge [] node {} (s7)
           ;
\end{tikzpicture}
\end{center}
%\caption{Realisation 1 of Example~\ref{ex:realisation}.}
\label{fig:tgb2}
\vspace{-4mm}
\end{subfigure}
    \begin{subfigure}[b]{0.37\textwidth}
\begin{center}
\begin{tikzpicture}[scale=0.9, every node/.style={transform shape}]
 \node[smallstate] (s1)                       {$100$};
 \node[smallstate]         (s2)  [right=of s1]   {$101$};
 \node[smallstate]         (s3)  at(0,-1.5)   {$110$};
 \node[smallstate]         (s4)  [right =of s3]  {$111$};

 \path     (s1)  edge [loop left] node {} (s1)
 		    (s2)  edge [] node {} (s4)
 		     (s2)  edge [] node {} (s1)
           (s3) edge [] node {} (s1)
           (s3)  edge [] node {} (s4)
           (s3)  edge [loop left] node {} (s3)
           (s4)  edge [loop right]       node {} (s4)
           ;
\end{tikzpicture}
\end{center}
%\caption{Realisation 2 of Example~\ref{ex:realisation}.}
\label{fig:tgb3}
\end{subfigure}
\caption{Transition graphs of two realisations in Example 2.}
\label{fig:realisationExample2}
\end{center}
\vspace{-6mm}
\end{figure}
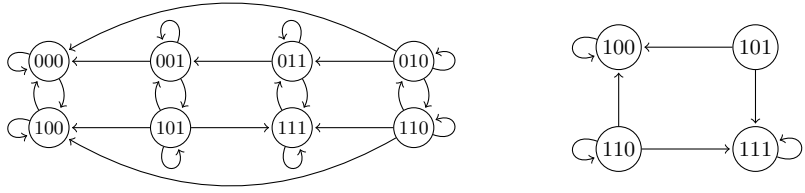

With Definition~\ref{def:preserve}, we have the following lemma and theorem. The proofs of all the
lemmas, theorems and corollaries in this paper are presented in Appendix~\ref{app:proofs}.
\begin{lemma}
\label{lemma:dec-attractor}
Given a~BN $G$ and an~elementary block $B$ in $G$, let $\Phi$ be the set of attractor states of
$G$ and $\Phi^{B}$ be the set of attractor states of $B$. If $B$ preserves the attractors of $G$,
then $\Phi \subseteq \mathcal{M}_G(\Phi^B)$.
%the mirror states of $B$'s attractor states contain all the attractor states of $M$.
\end{lemma}

\begin{theorem}
\label{theo:leavepres}
Given a~BN $G$, let $B$ be an~elementary block in $G$. $B$ preserves the attractors of $G$.
\end{theorem}

For an~elementary block $B$ in a~BN $G$, the mirror states of its attractor states cover all $G$'s
attractor states according to Lemma~\ref{lemma:dec-attractor} and Theorem~\ref{theo:leavepres}.
Therefore, by searching from the mirror states only instead of the whole state space, we can
detect all the attractor states of $G$.

We now proceed to consider the case of non-elementary blocks. For an~SCC $\Sigma_j$, if it has no
parent SCC, then this SCC can form an~elementary block; if it has at least one parent, then it
must have an~ancestor that has no parent, and all its ancestors $\Omega(\Sigma_j)$ together can
form an~elementary block, which is also a~BN. The SCC-based decomposition will result in at least
one elementary block and usually one or more non-elementary blocks. Moreover, for each
non-elementary block we can construct by merging all its predecessor blocks a~single parent
elementary block. We detect the attractors of the elementary blocks and use the detected
attractors to guide the values of the control nodes of their child blocks. The guidance is
achieved by considering \emph{realisations} of the dynamics of a~child block with respect to the
attractors of its parent elementary block.
%shortly referred to as realisations of the child block.
In some cases, a~realisation of a~block is
simply obtained by assigning new Boolean functions to the control nodes of the block. However, in
many cases, it is not this simple and a~realisation of a~block is obtained by explicitly
constructing a~transition system of this block corresponding to the considered attractor of the
elementary parent block. Since the parent block of a~non-elementary block may have more than one
attractor, a~block may have more than one realisation.

By the following two definitions, we explain in details what realisations are. We first introduce
the concept of crossability and cross operations in Definition~\ref{de:cross}. The concept of
crossability specifies a~special relation between states of a~non-elementary block and of its
parent blocks, while the cross operations are used for merging attractors of two blocks when
recovering the attractors of the original BN.

\begin{definition}[Crossability, Cross operations]
\label{de:cross}
Let $G$ be a~BN and let $B_i$ be a~non-elementary block in $G$ with the set of nodes
$V^{B_i}=\{v_{p_1},v_{p_2},\ldots,v_{p_s},v_{q_1},v_{q_2},\ldots,\allowbreak v_{q_t}\}$, where
$q_k$ ($k \in [1,t]$) are the indices of the control nodes also contained in $B_i$'s parent block
$B_j$ and $p_k$ ($k \in [1,s]$) are the indices of the remaining nodes. We denote the set of nodes
in $B_j$ as $V^{B_j}=\{v_{q_1},v_{q_2},\ldots,v_{q_t}, v_{r_1},v_{r_2},\ldots,v_{r_u}\}$, where
$r_k$ ($k \in [1,u]$) are the indices of the non-control nodes in $B_j$. Let further
$\boldsymbol{x}^{B_i} = (x_1,x_2,\ldots,x_s,y^i_1,y^i_2,\ldots,y^i_t)$ be a~state of $B_i$ and
$\boldsymbol{x}^{B_j} = (y^j_1,y^j_2,\ldots,y^j_t,z_1,z_2,\ldots,\allowbreak z_u)$ be a~state of
$B_j$. States $\boldsymbol{x}^{B_i}$ and $\boldsymbol{x}^{B_j}$ are said to be \emph{crossable},
denoted as $\boldsymbol{x}^{B_i} ~\mathcal{C} ~\boldsymbol{x}^{B_j}$, if the values of their
common nodes are the same, i.e., $y^i_k=y^j_k$ for all $k \in [1,t]$. The cross operation of two
crossable states $\boldsymbol{x}^{B_i}$ and $\boldsymbol{x}^{B_j}$ is defined as
$\Pi(\boldsymbol{x}^{B_i},\boldsymbol{x}^{B_j})=(x_1,x_2,\ldots,x_s,y^i_1,y^i_2,\ldots,y^i_t,
z_1,z_2,\ldots,z_u)$. The notion of crossability naturally extends to two elementary blocks;
%by definition any two states of two elementary blocks are crossable.
any two states of any two elementary blocks are always crossable.

We say a~set of states $S^{B_i} \subseteq X^{B_i}$ and a~set of states $S^{B_j} \subseteq X^{B_j}$
are crossable, denoted as $S^{B_i} ~\mathcal{C} ~S^{B_j}$, if at least one of the set is empty or
the following two conditions hold: 1) for any state $\boldsymbol{x}^{B_i} \in S^{B_i}$, there
always exists a~state $\boldsymbol{x}^{B_j} \in S^{B_j}$ such that $\boldsymbol{x}^{B_i}$ and
$\boldsymbol{x}^{B_j}$ are crossable; 2) vice versa. The cross operation of two crossable
non-empty sets of states $S^{B_i}$ and $S^{B_j}$ are defined as
$\Pi(S^{B_i},S^{B_j})=\{\Pi(\boldsymbol{x}^{B_i},\boldsymbol{x}^{B_j}) ~|~ \boldsymbol{x}^{B_i}
\in S^{B_i}, \boldsymbol{x}^{B_j} \in S^{B_j}$ and $\boldsymbol{x}^{B_i} ~\mathcal{C}
~\boldsymbol{x}^{B_j}\}$. When one of the two sets is empty, the cross operation simply returns
the other set, i.e., $\Pi(S^{B_i},S^{B_j})=S^{B_i}$ if $S^{B_j}=\emptyset$ and
$\Pi(S^{B_i},S^{B_j})=S^{B_j}$ if $S^{B_i}=\emptyset$.

Let $\mathcal{S}^{B_i}=\{S^{B_i} ~|~ S^{B_i}\subseteq X^{B_i}\}$ be a~set of states set in $B_i$
and $\mathcal{S}^{B_j}=\{S^{B_j} ~|~ S^{B_j}\subseteq X^{B_j}\}$ be a~set of states set in $B_j$.
We say $\mathcal{S}^{B_i}$ and $\mathcal{S}^{B_j}$ are crossable, denoted as $\mathcal{S}^{B_i}
~\mathcal{C}~\mathcal{S}^{B_j}$ if for any states set $S^{B_i} \in \mathcal{S}^{B_i}$, there
always exists a~states set $S^{B_j} \in \mathcal{S}^{B_j}$ such that $S^{B_i}$ and $S^{B_j}$ are
crossable; 2) vice versa. The cross operation of two crossable sets of states sets
$\mathcal{S}^{B_i}$ and $\mathcal{S}^{B_j}$ are defined as
$\Pi(\mathcal{S}^{B_i},\mathcal{S}^{B_j})=\{\Pi(S_i,S_j) ~|~ S_i \in \mathcal{S}^{B_i}, S_j \in
\mathcal{S}^{B_j}$ and $S_i ~\mathcal{C} ~S_j\}$.
\end{definition}

%\begin{proposition} \label{prop:crossable}
%Let $V^C$ be the set of control nodes shared by two blocks $B_i$ and
%$B_j$, i.e., $V^C= V^{B_i} \cap V^{B_j}$, and
%$S^{B_i} \subseteq X^{B_i}$ and $S^{B_j} \subseteq X^{B_j}$. The
%crossable relation $S^{B_i} ~\mathcal{C} ~S^{B_j}$ is equivalent to
%$\pi_{C}(S^{B_i}) = \pi_{C}(S^{B_j})$.
%\end{proposition}

With the crossability defined, the definition of a~realisation is now given as follows.

\begin{definition}[Realisation of a~block]
\label{def:realisation}
Let $B_i$ be a~non-elementary block formed by merging an~SCC with its control nodes. Let nodes
$u_1, u_2, \ldots, u_r$ be all the control nodes of $B_i$ which are also contained by its single
and elementary parent block $B_j$ (we can always merge all $B_i$'s ancestor blocks to form $B_j$
if $B_i$ has more than one parent block or has a~non-elementary parent block).
Let $A^{B_j}_1, A^{B_j}_2,\ldots, A^{B_j}_t$ be the AS' of $B_j$. For any $k \in [1,t]$,
a~\emph{realisation} of block $B_i$ with respect to $A^{B_j}_k$ is a~state transition system
such that
\begin{enumerate}
\item a~state of the system is a~vector of the values of all the nodes in the block;
\item the state space of this realisation is crossable with $A^{B_j}_k$;
%For any $k \in [1,t]$, block $B_i$ has a \emph{realisation} of $a_{jk}$ by
%regulating the control nodes in the same way as in attractor $a_{jk}$:
%1) the state space of this realisation is crossable with $a_{jk}$;
\item for any transition $\boldsymbol{x}^{B_i} \rightarrow \boldsymbol{\tilde{x}}^{B_i}$ in this
  realisation, if this transition is caused by a~non-control node, the transition should be
  regulated by the Boolean function of this node; if this transition is caused by the updating
  of a~control node, one can always find two states $\boldsymbol{x}^{B_j}$ and
  $\boldsymbol{\tilde{x}}^{B_j}$ in $A^{B_j}_k$ such that there is a~transition from
  $\boldsymbol{x}^{B_j}$ to $\boldsymbol{\tilde{x}}^{B_j}$ in $A^{B_j}_k$,
  $\boldsymbol{x}^{B_i}~\mathcal{C}~\boldsymbol{x}^{B_j}$ and
  $\boldsymbol{\tilde{x}}^{B_i}~\mathcal{C}~\boldsymbol{\tilde{x}}^{B_j}$;
\item for any transition $\boldsymbol{x}^{B_j} \rightarrow \boldsymbol{\tilde{x}}^{B_j}$ in
  $A^{B_j}_k$, one can always find a~transition $\boldsymbol{x}^{B_i} \rightarrow
  \boldsymbol{\tilde{x}}^{B_i}$ in this realisation such that
  $\boldsymbol{x}^{B_i}~\mathcal{C}~\boldsymbol{x}^{B_j}$ and
  $\boldsymbol{\tilde{x}}^{B_i}~\mathcal{C}~\boldsymbol{\tilde{x}}^{B_j}$.
\end{enumerate}
\end{definition}

Constructing realisations for a~non-elementary block is the key process for obtaining its
attractors. For each realisation, the construction process requires the knowledge of all the
transitions in the corresponding attractor of the parent block.
In Section~\ref{sec:algorithms}, we explain in details how to implement it with BDDs.
%We then perform an~OR operation of this BDD with the BDD
%encoding the transition relations containing the nodes in the SCC of the block to construct the
%transition relations of the realisation.

\begin{example}
\label{ex:realisation}
Consider the BN shown in Figure~\ref{fig:sccde}.
%The Boolean functions of this network are given as follows:
%\begin{equation}
%\begin{cases}
%  \begin{tabular}{p{2.2cm}p{2.4cm}p{3.1cm}p{2.5cm}}
%  $f_1=x_1 \wedge x_2$,  & $f_2=x_1 \vee \neg x_2$, &$ f_3=\neg x_4$,
%&$ f_4=x_1 \wedge \neg x_3$,  \\ $f_5=x_2 \wedge x_6$,  &$f_6=x_5$,
% &$f_7=(x_1 \vee x_6) \wedge x_8$,& $f_8=x_7 \vee  x_8.$
%  \end{tabular}
%\end{cases}
%\end{equation}
The network contains four SCCs $\Sigma_1, \Sigma_2, \Sigma_3$ and $\Sigma_4$. For any $\Sigma_i$
($i \in [1,4]$), we form a~block $B_i$ by merging $\Sigma_i$ with its control nodes. Block $B_1$
is an~elementary block and its transition graph is shown in Figure~\ref{fig:tgb}.
Block $B_1$ has two attractors, i.e., $\{(0*)\}$ and
$\{(11)\}$. Regarding the first attractor, block $B_3$ has a~realisation by setting node $v_2$ to
contain the following transitions $\{(0)\rightarrow(*), (1) \rightarrow (*)\}$.
The transition graph of this realisation is shown in Figure~\ref{fig:realisationExample2} (left).
Regarding the second attractor, block $B_3$
has a~realisation by setting node $v_2$ to contain only the transition $\{(1)\rightarrow(1)\}$.
Its transition graph is shown in Figure~\ref{fig:realisationExample2} (right).
\end{example}

A~realisation of a~block takes care of the dynamics of the undetermined nodes and instantiates
a~transition system of the block. Therefore, we can extend the attractor definition to
realisations and to non-elementary blocks as follows.
%assigns initial values and Boolean functions to undetermined nodes.
%After realisation, there are no undetermined nodes in a block.
%Therefore, a realisation of a non-elementary block can be viewed as a BN with restricted initial states.
\begin{definition}[Attractors of a~non-elementary block]
An~attractor of a~realisation of a~non-elementary block is a~set of states satisfying that any
state in this set can be reached from any other state in this set and no state in this set can
reach any other state that is not in this set. The attractors of a~non-elementary block is the set
of the attractors of all realisations of the block.
\end{definition}

With the definition of attractors of non-elementary blocks, we can relax
Definition~\ref{def:realisation} by allowing $B_j$ to be a~single and either elementary or
non-elementary parent block with known attractors. This is due to the fact that when forming the
realisations of a~non-elementary block, we only need the attractors of its parent block that
contains all its control nodes, no matter whether this parent block is
elementary or not. In other words, computing attractors for non-elementary blocks requires the
knowledge of the attractors of its parent block that contains all its control nodes. Therefore, we
need to consider blocks in a~specific order which guarantees that when computing attractors for
block $B_i$, the attractors of its parent block that contains all $B_i$'s control nodes are
already available. To facilitate this, we introduce the concept of a~credit as follows.

\begin{definition}[Credit]
\label{def:credit}
Given a~BN $G$, an~elementary block $B_i$ of $G$ has a~credit of 0, denoted as
$\mathcal{P}(B_i)=0$. Let $B_j$ be a~non-elementary block and $B_{j_1},\ldots,B_{j_{p(j)}}$ be all
its parent blocks. The credit of $B_j$ is defined as
$\mathcal{P}(B_j)=max_{k=1}^{p(j)}(\mathcal{P}(B_{j_k}))+1$.
\end{definition}

%===============================================
\subsection{Recover attractors of the original BN}
\label{ssec:recoverattractor}
%===============================================

After computing attractors for all the blocks, we need to recover attractors for the original BN.
This is achievable by the following theorem. %for recovering the attractors of two blocks.

\begin{theorem}
\label{theorem:recover}
Given a~BN $G$ with $B_i$ and $B_j$ being its two blocks, let $\mathcal{A}^{B_i}$ and
$\mathcal{A}^{B_j}$ be the set of attractors for $B_i$ and $B_j$, respectively. Let $B_{i,j}$ be
the block got by merging the nodes in $B_i$ and $B_j$. If $B_i$ and $B_j$ are both elementary
blocks or $B_i$ is an~elementary and single parent block of $B_j$, then
$\mathcal{A}^{B_i}~\mathcal{C}~\mathcal{A}^{B_j}$ and $\Pi(\mathcal{A}^{B_i},\mathcal{A}^{B_j})$
is the set of attractors of $B_{i,j}$.
\end{theorem}

Finally, from Theorem~\ref{theorem:recover} we obtain the following corollary which states that
for specific configurations of blocks, certain orderings according to which the blocks are merged
are equivalent in terms of the resulting attractor set for the merged block.
\begin{corollary}
\label{corollary:order}
Given a~BN $G$ with $B_i$, $B_j$, and $B_k$ being its three blocks, let $\mathcal{A}^{B_i}$,
$\mathcal{A}^{B_j}$, and $\mathcal{A}^{B_k}$ be the sets of attractors for blocks $B_i$, $B_j$,
and $B_k$, respectively. If the three blocks are all elementary blocks or $B_i$ is an~elementary
block and it is the only parent block of $B_j$ and $B_k$, it holds that
$\Pi(\Pi(\mathcal{A}^{B_i},\mathcal{A}^{B_j}),\mathcal{A}^{B_k})=
 \Pi(\Pi(\mathcal{A}^{B_i},\mathcal{A}^{B_k}),\mathcal{A}^{B_j})$.
\end{corollary}

The above developed theoretical background with Theorem~\ref{theorem:recover} being its core
result, allows us to design a~new decomposition-based approach towards detection of attractors
in large asynchronous BNs. The idea is as follows. We divide a~BN into blocks according to the
detected SCCs. We order the blocks in the ascending order based on their credits and detect
attractors of the ordered blocks one by one in an~iterative way. According to
Theorem~\ref{theorem:recover}, we can perform a~cross operation for any two elementary blocks
(credits 0) or an~elementary block (credit 0) with one of its child blocks (credit~1) which has no
other parent blocks to recover the attractors of the two merged blocks. The resulting merged block
will form a~new elementary block, i.e., one with credit 0. By iteratively performing the cross
operation until a~single elementary block containing all the nodes of the BN is obtained, we can
recover the attractors of the original BN. The details of this new approach are discussed in the
next section.

%===============================================
\section{Implementation}
\label{sec:algorithms}
%===============================================
%In this section, we introduce the algorithms we use to detect attractors in asynchronous BNs.
%As mentioned before, encoding BNs in BDDs has the advantage of saving memory.
%Hence, our attractor detection algorithms are also based on BDD representations.
We first introduce a BDD-based algorithm to detect attractors for relatively small BNs. %in Section~\ref{ssec:bddalg}.
Then we describe how our SCC-based decomposition method can be implemented
using the BDD-based algorithm.

%%===============================================
\subsection{BDD-based attractor detection algorithm}
\label{ssec:bddalg}
%%===============================================

Attractors of an asynchronous BN are in fact bottom strongly connected components (BSCCs)
in the transition system of the BN. Thus, detecting attractors is the same as detecting the BSCCs.
Formally, the definition of BSCCs is given as follows.

\begin{definition}
A~bottom strongly connected component (BSCC) is an~SCC $\Sigma$ such that no state outside
$\Sigma$ is reachable from $\Sigma$.
\end{definition}

%Similarly to the concept of an attractor system, we refer to a~BSCC with the edges between the vertices in the
%BSCC as a~\emph{bottom strongly connected component system} (BSCCS).

We encode a~BN with BDDs,
and adapt the hybrid Tarjan algorithm described in Algorithm 7 of~\cite{KPQ11} to detect BSCCs
in the corresponding transition system of the BN.
%% This hybrid Tarjan's algorithm is the adaptation of a variant of Tarjan's algorithm~\cite{NUUTILA19949} on the BDD structure.
%% Then for each of the detected SCCs, we verify whether it is a~BSCC.
Given a transition system $\mathcal{T}=\langle S,S_0,\allowbreak
T\rangle$, %% and a set of states $X\subseteq S$ such that $X={\it Image}(X, T)$
our attractor detection algorithm {\sc Detect}$(\mathcal{T})$ in Algorithm~\ref{alg:detect}
computes the set of BSCCs in $\mathcal{T}$. If $\mathcal{T}$ is
converted from a BN $G$, then {\sc Detect}$(\mathcal{T})$ computes all the
attractors of $G$.
The correctness of Algorithm~\ref{alg:detect} is guaranteed by the following two propositions.

\begin{proposition} \label{theo:firstscc}
The first SCC returned by the Tarjan's algorithm is a BSCC.
\end{proposition}

\begin{proposition} \label{theo:newpred}
If a state that reaches a~BSCC is located outside the BSCC, then this state is not contained by any BSCC.
\end{proposition}

The first proposition can be deduced from the fact that the Tarjan's
algorithm is a~depth-first search. The second one comes from the
definition of BSCCs, as no states inside a~BSCC can lead to a~state in
any other BSCC. In Algorithm~\ref{alg:detect}, the hybrid Tarjan
algorithm $HybridTarjan(s, T)$ takes as input a starting state $s$ and
the transition relation $T$. When it finds the first SCC $\Sigma$
(also a~BSCC), which is reached from $s$, it terminates immediately
and returns $\Sigma$.

With the use of BDD representation, {\sc Detect}$(\mathcal{T})$ can deal with relatively small BNs
(e.g., a~BN with tens of nodes) with small memory usage.
Moreover, the computation of SCCs can also benefit from the efficient BDD operations.
However, real life biological BNs usually contain hundreds of nodes
and the state space is exponential in the number of nodes.
Therefore, {\sc Detect}$(\mathcal{T})$ would still suffer from the state space explosion problem
when dealing with large BNs. Thus for large BNs,
we propose to use the SCC-based decomposition method as described in Section~\ref{sec:decomposition}.

\begin{algorithm}[p]
\caption{Attractor detection using the hybrid Tarjan's algorithm}
\label{alg:detect}
\vspace{-5mm}
\begin{multicols}{2}
\begin{algorithmic}[1]
\Procedure{Detect}{$\mathcal{T}$}
\State  $\mathcal{A} := \emptyset$; $X := S$; \hfill{\it //$S$ is from $\mathcal{T}$}
\While {$X\not= \emptyset$}
\State Randomly pick a state $s\in X$;
\State $\Sigma := HybridTarjan(s, T)$;
\State $\mathcal{A} := \mathcal{A}\cup \Sigma$;
\State $X := X\backslash {\it Predecessors}(\Sigma, T)$;
\EndWhile
\State \Return $\mathcal{A}$.
\EndProcedure
\end{algorithmic}
\end{multicols}
\vspace{-2mm}
\end{algorithm}

\begin{algorithm}[p]
\caption{SCC-based decomposition algorithm}
\label{alg:scc}
\begin{algorithmic}[1]
\Procedure{SCC\_Detect}{$G$, $\mathcal{T}$}
\State $B:=~${\sc Form\_Block}($G$);  $\mathcal{A}:=\emptyset;~B_a:=\emptyset;~k:=$ size of $B$;
\label{line:blocks}
\State initialise dictionary $\mathcal{A}^{\ell}$; \hfill{\it //$\mathcal{A}^{\ell}$ is
a~dictionary storing the set of attractors for each block}
\For {$i:=1;i<=k;i++$}
\label{line:forloopstart}
\If{$B_i$ is an~elementary block}
\label{line:detectblockstart}
\State $\mathcal{T}^{B_i}:=$ transition system converted from $B_i$;  \hfill{\it //see Section~\ref{ssec:encoding} for more details}
\State $\mathcal{A}_i:=~${\sc Detect}$(\mathcal{T}^{B_i})$;
\Else {  $\mathcal{A}_i:=\emptyset$;}
\If{$B_i^{p}$ is the only parent block of $B_i$}
\State  $\mathcal{A}_i^{p}:=\mathcal{A}^{\ell}.getAtt(B_i^{p})$;
\hfill{\it //obtain attractors of $B_i^p$}
%\Else { $T^{B_c}:=$ transition relation of $B_c$;~$\mathcal{A}_i^{p}:=~${\sc Detect}$(T^{B_c},\Phi_c);$}
\Else {~$B^p:=\{B_1^{p},B_2^{p},\ldots,B_m^{p}\}$ be the ancestor blocks of $B_i$ (ascending ordered); }
\State $ B_c:= B_1^p;$ \hfill{\it //$B^p$  is ordered based on credit}
\label{line:detectmerge}
\For {$j:=2; j<= m; j++$}
\label{line:mergeforloopstart}
\State $B_{c,j}:=$ a~new block containing nodes in $B_c$ and $B_j^{p}$;
\If {$(\mathcal{A}_i^p:=\mathcal{A}^{\ell}.getAtt(B_{c,j}))==\emptyset$}
\State $\mathcal{A}_i^p:=\Pi(\mathcal{A}^{\ell}.getAtt(B_c),\mathcal{A}^{\ell}.getAtt(B_j))$;
 $\mathcal{A}^{\ell}.add(B_{c,j},\mathcal{A}^p_i)$
\label{line:api}
\EndIf
\State $B_c:=B_{c,j}$; %\hfill{\it //merge $B_c$ with $B_j$}
\EndFor
\label{line:mergeforloopstop}
\EndIf
\For {$A \in \mathcal{A}_i^{p}$} %{$T:= T(A);$} \hfill{\it //$T(A)$ obtains the transition relation for $A$}
\label{line:start}
\State $\mathcal{T}^{B_i}(A):=\langle S^{B_i}(A), T^{B_i}(A) \rangle$; \hfill{\it //obtain the realisation of $B_i$ with $A$}
\State $\mathcal{A}_i:=\mathcal{A}_i\cup~${\sc Detect}$(\mathcal{T}^{B_i}(A))$;

\label{line:realisation}
\EndFor
\label{line:end}
\EndIf
\label{line:detectblockend}
\State $\mathcal{A}^{\ell}.add(B_{i},\mathcal{A}_i);$  \hfill{\it //the add operation will not add duplicated elements}
%\hfill{\it //$F(\mathcal{A}_i)$ returns all the states in attractors $\mathcal{A}_i$}
%\State $B_c:=$ merge $B_c$ with $B_i$;~$\Phi_c:=\Pi(\Phi_i,\Phi_c)$;
\If {$B_a!=\emptyset$}  ~$\mathcal{A}=\Pi(\mathcal{A}_i,\mathcal{A});$~$B_a:=B_{a,i};~\mathcal{A}^{\ell}.add(B_a,\mathcal{A});$
\Else {~$B_a:=B_i$}
\EndIf
 \label{line:cross}
\EndFor
\label{line:forloopend}
%\State $T:=$ transition relation of $G$; ~$\mathcal{A} = ~${\sc Detect}$(G,\Phi_c)$;~
\State \Return $\mathcal{A}$.
\EndProcedure \medskip
\Procedure{Form\_Block}{$G$}
\State decompose $G$ into SCCs and form blocks with SCCs and their control nodes;
\State order the blocks in an ascending order according to their credits;~$B:=(B_1,\ldots,B_k)$;
\State \Return $B$. \hfill {\it //$B$ is the list of blocks after ordering}
\EndProcedure
\end{algorithmic}
\end{algorithm}
%
%===============================================
\subsection{SCC-based decomposition algorithm}
\label{ssec:sccalg}
%===============================================

We describe the detection process in Algorithm~\ref{alg:scc}. This algorithm takes a~BN $G$ and its corresponding transition system $\mathcal{T}$ as inputs and outputs the set of attractors of
$G$. Lines~\ref{line:start}-\ref{line:end} of this algorithm describe the process for detecting
attractors of a~non-elementary block. The algorithm detects the attractors of all the realisations
of the non-elementary block and performs the union operation on the sets of detected attractors.
For this, if the non-elementary block has only one parent block, its attractors are already
computed as the blocks are considered in the ascending order with respect to their credits by the
main \textbf{for} loop in Line~\ref{line:forloopstart}. Otherwise, all the ancestor blocks are
considered in the \textbf{for} loop in
Lines~\ref{line:mergeforloopstart}-\ref{line:mergeforloopstop}. By iteratively applying the cross
operation in Line~\ref{line:api} to the attractor sets of the ancestor blocks in the ascending
order, the attractors of a~new block formed by merging all the ancestor blocks are computed as
assured by Theorem~\ref{theorem:recover}. The new block is in fact an~elementary block which is
a~single parent of the considered non-elementary block. By considering blocks in the
ascending order, the order in which blocks with the same credit are considered does not influence
the final result due to Corollary~\ref{corollary:order}.
The correctness of the algorithm is stated as Theorem~\ref{theo:scc-identify}.
%The correctness of the algorithm is
%stated as the following theorem which proof is given in Appendix~\ref{app:proofs}.

\begin{theorem}
\label{theo:scc-identify}
Algorithm~\ref{alg:scc} correctly identifies the set of attractors of a~given BN $G$.
\end{theorem}

The algorithm stores all computed attractors for the original SCC
blocks and all auxiliary merged blocks in the dictionary structure
$\mathcal{A}^{\ell}$. We use BDDs to encode transitions and the
realisations are performed via BDD operations directly. Given a~BN
$G(V, \boldsymbol{f})$ with $n$ nodes, our implementation, which is
based on the CUDD library~\cite{CUDD}, encodes the whole network with
$2n$ BDD variables. Each state in $G$ is encoded by $n$ BDD variables,
and a~projection of a~state on a~subset of nodes $V'\subseteq V$ is
performed by setting all BDD variables for nodes in $V\backslash V'$
to ``-'', which represents that its value can be either 0 or 1, and
therefore, can be ignored. As a~state for a~block $B$ is encoded by
$|V^B|$ BDD variables, the variables in $V\backslash V^B$ are set to
``-'' in the BDD representation. This way, after we verify that $S^{B_i}$
and $S^{B_j}$ are crossable, i.e.,$S^{B_i} ~\mathcal{C} ~S^{B_j}$,
%using Proposition~\ref{prop:crossable},
the cross operation $\Pi(S^{B_i},S^{B_j})$ is equivalent to the AND
operation on two BDDs, i.e., $bdd_{S^{B_i}}$ and $bdd_{S^{B_j}}$ encoding
$S^{B_i}$ and $S^{B_j}$, respectively. Formally, we have that
$\Pi(S^{B_i},S^{B_j})=bdd_{S^{B_i}}\cap bdd_{S^{B_j}}$. Let
$\mathcal{T}^B=\langle S^B, T^B\rangle$ be the transition system
converted from block $B$, and let $V^C$ be the set of control nodes in
$B$. The set of states $S^B(A)$ of the realisation of block $B$ with
respect to attractor $A$ is $\mathcal{M}_{B}(\pi_C(A))$ and the transition
relation $T^{B}(A)$ of the realisation is $T^B|_{S^B(A)}$.
%% \JP{Hongyang, please explain the BDD
%%   implementations of $T$ (line 21), ${\cal R}$ (line 22), and $\Pi$
%%   (line 27).}
We continue to illustrate in Example~\ref{ex:merge} how
Algorithm~\ref{alg:scc} detects attractors.

\begin{figure}[t!]
%    \begin{subfigure}[b]{0.37\textwidth}
%\begin{center}
%\begin{tikzpicture}[scale=0.9, every node/.style={transform shape}]
% \node[smallstate] (s1)                       {$100$};
% \node[smallstate]         (s2)  [right=of s1]   {$101$};
% \node[smallstate]         (s3)  at(0,-1.2)   {$110$};
% \node[smallstate]         (s4)  [right =of s3]  {$111$};
%
% \path     (s1)  edge [loop left] node {} (s1)
% 		    (s2)  edge [loop right] node {} (s2)
% 		     (s2)  edge [] node {} (s1)
%           (s3) edge [] node {} (s1)
%           (s3)  edge [] node {} (s4)
%           (s3)  edge [loop left] node {} (s3)
%           (s4)  edge [loop right]       node {} (s4)
%           ;
%\end{tikzpicture}
%\end{center}
%\caption{Realisation 2 of Example~\ref{ex:realisation}.}
%\label{fig:tgb3}
%\end{subfigure}
   \begin{subfigure}[b]{0.48\textwidth}
    \begin{center}
\begin{tikzpicture}[scale=0.9, every node/.style={transform shape}]
 \node[smallstate] (s1)                       {$000$};
 \node[smallstate]         (s2)  [right=of s1]   {$010$};
 \node[smallstate]         (s3)  at(0,-1.2) {$001$};
 \node[smallstate]         (s4)  at(2,-1.2) {$011$};

 \path     (s1)  edge [loop left] node {} (s1)
  (s1)  edge [] node {} (s2)
   (s2)  edge [loop right] node {} (s2)
   (s3)  edge [loop left] node {} (s3)
   (s3)  edge [] node {} (s1)
   (s4)  edge [] node {} (s2)
   (s4)  edge [] node {} (s3)
   (s4)  edge [loop right] node {} (s4)
           ;
\end{tikzpicture}
\end{center}
%\caption{Realisation 1 of $B_2$.}
%\label{fig:tgb4}
\end{subfigure}
    \begin{subfigure}[b]{0.48\textwidth}
\begin{center}
\begin{tikzpicture}[scale=0.9, every node/.style={transform shape}]
 \node[smallstate] (s1)                       {$100$};
 \node[smallstate]         (s2)  [right=of s1]   {$101$};
 \node[smallstate]         (s3)  at(0,-1.2)   {$110$};
 \node[smallstate]         (s4)  [right =of s3]  {$111$};

 \path     (s1)  edge [loop left] node {} (s1)
 			  (s1) edge [] node {} (s3)
 			  (s1) edge [] node {} (s2)
 		    (s2)  edge [loop right] node {} (s2)
           (s3)  edge [loop left] node {} (s3)
           (s4)  edge [loop right]       node {} (s4)
           (s4) edge [] node {} (s2)
           (s4) edge [] node {} (s3)
           ;
\end{tikzpicture}
\end{center}
%\caption{Realisation 2 of $B_2$.}
%\label{fig:tgb5}
\end{subfigure}
\caption{Transition graphs of the two realisations for block $B_2$.}
\label{fig:tgs}
\vspace{-3mm}
\end{figure}
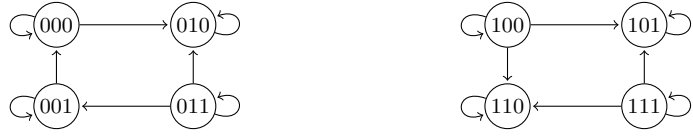

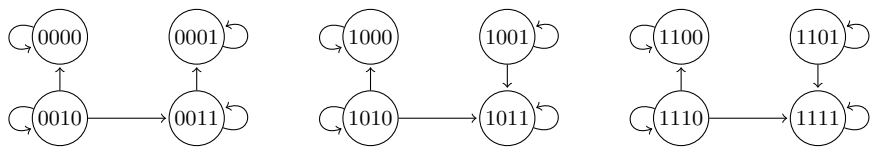
\begin{figure}[!t]
    \begin{subfigure}[b]{0.33\textwidth}
    \begin{center}
\begin{tikzpicture}[scale=0.9, every node/.style={transform shape}]
 \node[smallstate] (s1)                       {$0000$};
 \node[smallstate]         (s2)  at(0,-1.2)  {$0010$};
 \node[smallstate]         (s3)  [right=of s1]  {$0001$};
 \node[smallstate]         (s4)  at(2,-1.2)  {$0011$};

 \path     (s1)  edge [loop left] node {} (s1)
  (s2)  edge [] node {} (s1)
  (s2)  edge [] node {} (s4)
   (s2)  edge [loop left] node {} (s2)
   (s3)  edge [loop right] node {} (s3)
   (s4)  edge [] node {} (s3)
   (s4)  edge [loop right] node {} (s4)
           ;
\end{tikzpicture}
\end{center}
\end{subfigure}
    \begin{subfigure}[b]{0.33\textwidth}
\begin{center}
\begin{tikzpicture}[scale=0.9, every node/.style={transform shape}]
 \node[smallstate] (s1)                       {$1000$};
 \node[smallstate]         (s2)  [right=of s1]   {$1001$};
 \node[smallstate]         (s3)  at(0,-1.2)   {$1010$};
 \node[smallstate]         (s4)  [right =of s3]  {$1011$};

 \path     (s1)  edge [loop left] node {} (s1)
 		    (s2)  edge [loop right] node {} (s2)
 		    (s2) edge [] node {} (s4)
           (s3)  edge [loop left] node {} (s3)
           (s3) edge [] node {} (s1)
           (s3) edge [] node {} (s4)
           (s4)  edge [loop right]       node {} (s4)
           ;
\end{tikzpicture}
\end{center}
\end{subfigure}
    \begin{subfigure}[b]{0.33\textwidth}
\begin{center}
\begin{tikzpicture}[scale=0.9, every node/.style={transform shape}]
 \node[smallstate] (s1)                       {$1100$};
 \node[smallstate]         (s2)  [right=of s1]   {$1101$};
 \node[smallstate]         (s3)  at(0,-1.2)  {$1110$};
 \node[smallstate]         (s4)  [right =of s3]  {$1111$};

 \path     (s1)  edge [loop left] node {} (s1)
 		    (s2)  edge [loop right] node {} (s2)
 		    (s2) edge [] node {} (s4)
           (s3)  edge [loop left] node {} (s3)
           (s3) edge [] node {} (s1)
           (s3) edge [] node {} (s4)
           (s4)  edge [loop right]       node {} (s4)
           ;
\end{tikzpicture}
\end{center}
\end{subfigure}
\caption{Transition graphs of the three realisations for block $B_4$.}
\label{fig:tgs3}
\vspace{-3mm}
\end{figure}

\begin{example}
\label{ex:merge}
Consider the BN  shown in Example~\ref{ex:realisation} and its four blocks. Block $B_1$ is
an~elementary block and it has two attractors, i.e., $\mathcal{A}_1=\{\{(0*)\},\{(11)\}\}$.
To detect the attractors of block $B_2$, we first form realisations of $B_2$ with respect to the
attractors of its parent block $B_1$.
$B_1$ has two attractors so there are two realisations for $B_2$. The transition graphs of the two realisations are shown in Figure~\ref{fig:tgs}.
We get three attractors for block $B_2$, i.e., $\mathcal{A}_2=\{\{(010)\},\{(101)\},\{(110)\}\}$. Performing a~cross operation,
we get the attractors of the merged block $B_{1,2}$, i.e., $\mathcal{A}_{1,2}=\Pi(\mathcal{A}_1,\mathcal{A}_2)=\{\{(0*10)\},\allowbreak
\{(1101)\},\{(111\allowbreak0)\}\}$. In Example~\ref{ex:realisation},
we have shown the two realisations of $B_3$ with respect to the two attractors of $B_1$. Clearly,
$B_3$ has three attractors, i.e.,
$\mathcal{A}_3=\{\{(*00)\},\{(100)\},\{(111)\}\}$. Merging $B_{1,2}$ and $B_3$,
we get the attractors of the merged block $B_{1,2,3}$, i.e., $\mathcal{A}_{1,2,3}=\Pi(\mathcal{A}_{1,2},\mathcal{A}_3)=
\{\{(0*1000)\},\{(110100)\},\{(11\allowbreak0111)\},\{(1\allowbreak11000)\},\{(111\allowbreak011)\}\}$.
$B_4$ has two parent blocks. Therefore, we need to merge $B_4$'s ancestors ($B_1$ and $B_3$) as
its new parent block. After merging, we get the attractors of the merged block as
$\mathcal{A}_{1,3}=\Pi(\mathcal{A}_1,\mathcal{A}_3)=\{\{(0*00)\},\{(1100)\},\{(1111)\}\}.$
There are three attractors so there will be three realisations for block $B_4$. The transition
graphs of the three realisations are shown in Figure~\ref{fig:tgs3}. From the transition graphs,
we easily get the attractors of $B_4$, i.e.,
$\mathcal{A}_4=\{\{(0000)\},\{(0001)\},\{(1000)\},\{(1011)\},\allowbreak\{(1100)\},\{(1111)\}\}$.
Now the attractors for all the blocks have been detected. We can then obtain the attractors of
the BN by applying one more cross operation, i.e.,
$\mathcal{A}=\mathcal{A}_{1,2,3,4}=\Pi(\mathcal{A}_{1,2,3},\mathcal{A}_4)=
\{\{(0*100000)\},\{(0*100001)\},\{(11010000)\},\{(1101\allowbreak0011)\},\{(11011100)\},
\{(11011111)\},\{(11100000)\},\{(11100011)\},\{(11101100)\},\allowbreak\{\allowbreak(11101111)\}\}$.
\end{example}

%\input{optimization}

%===============================================
\section{Evaluation}
\label{sec:eva}
%===============================================

We have implemented the decomposition algorithm presented in Section~\ref{sec:algorithms} in the
model checker MCMAS~\cite{MCMAS}. In this section, we demonstrate the efficiency of our
method using two real-life biological systems. One is a~logical MAPK network model
of~\cite{GCT13} containing 53 nodes and the other is a~Boolean network model of apoptosis,
originally presented in~\cite{SSVSSBEMS09}, containing 97 nodes. All the experiments are conducted
on a~computer with an~Intel Xeon W3520@2.67GHz CPU and 12GB memory. Notice that we tried to apply
genYsis~\cite{ALG07} to these two systems, but it failed in both cases to detect attractors within
5 hours.
%
%Note that we also try to compute the attractors for these two models with the tool genYsis~\cite{ALG07}.
%However, genYsis failed to detect attractors within 5 hours.

\smallskip
\noindent \textbf{MAPK network.}
Mitogen-activated protein kinases (MAPKs) are a~family of serine/ threonine kinases that transduce
biochemical signals from the cell membrane to the nucleus in response to a~wide range of stimuli.
%
%such as growth factors, hormones, inflammatory cytokines and environmental stresses.
%Cascades of these kinases participate in multiple intracellular signalling pathways that control a~wide range
%of cellular processes, e.g. cell cycle machinery, differentiation, survival and apoptosis. MAPK
%pathways are highly evolutionary conserved among all eukaryotic cells and allow the cells to
%respond coordinately to multiple and diverse inputs. To date, three main pathways have been
%extensively studied: Extracellular Regulated Kinases (ERK), Jun $\textrm{NH}_2$ Terminal Kinases
%(JNK), and p38 Kinases (p38), named after their specific MAPK kinases involved. These pathways are
%characterised by enormous crass-talk with each other, which gives rise to a~complex network of
%molecular interactions (\cite{KN08}). Malfunctioning of MAPK signalling mechanisms is often
%observed in cancer~\cite{DHRK07}. Therefore, a~deeper comprehension of the MAPK pathways and their
%interactions is of utter importance to elucidate the roles of MAPKs in the development and
%progression of cancer. This in turn is crucial for the development of new, effective therapeutic strategies.
%
In~\cite{GCT13} a~predictive dynamical Boolean model of the MAPK network is presented. It
recapitulates observed responses of the MAPK network to characteristic stimuli in selected
urinary bladder cancers together with its specific contribution to cell fate decision on
proliferation, apoptosis, and growth arrest. The wiring of the logical model
of~\cite{GCT13} is shown in Figure~\ref{fig:mapk_structure} in Appendix~\ref{app:mapk}.
In our study we consider two mutants of the model: one with EGFR over-expression and the other
with FGFR3 activating mutation which correspond to the r3 and r4 variants of~\cite{GCT13},
respectively, and therefore we refer to them as as MAPK\_r3 and MAPK\_r4. However, in contrast to
the original variants r3 and r4, we do not set the values for the four stimuli nodes to 0 but
perform the computations for all $2^4$ possible fixed sets of values for these nodes. For the
remaining nodes, all possible initial states are considered as in~\cite{GCT13}.
In consequence, our results for MAPK\_r3 and MAPK\_r4 include the attractors for variants r7, r13 and r8, r14
of~\cite{GCT13}, respectively.
We compute the attractors of the MAPK\_r3 and MAPK\_r4 BNs using
both the BDD-based algorithm, i.e., Algorithm~\ref{alg:detect} and our decomposition
algorithm, i.e., Algorithm~\ref{alg:scc}. We show in the left part of Table~\ref{tab:evaluation}
the number of attractors and the computational time costs for both mutants.
Besides, we show the speedups of Algorithm~\ref{alg:scc} with respect to Algorithm~\ref{alg:detect}.
Notice that our computations are performed for the full model presented in
Figure~\ref{fig:mapk_structure} contrary to~\cite{GCT13}, where various reduced models were used
for the computations of attractors.

\smallskip
\noindent \textbf{Apoptosis network.}
Apoptosis is a~process of programmed cell death and has been linked to many diseases.
It is often regulated by several signaling pathways extensively linked by crosstalks.
We take the apoptosis signalling network presented in~\cite{SSVSSBEMS09} and recast it into the
Boolean network framework: a~BN model which compromise 97 nodes. In this network, there are
10 input nodes. One of them is a~housekeeping node which value is fixed to 1 and which is used to
model constitutive activation of certain nodes in the network. For the wiring of the BN model, see
Figure~\ref{fig:structure} in Appendix~\ref{app:apoptosis}. Similar to the MAPK network, we
compute the attractors of the apoptosis network with both Algorithm~\ref{alg:detect} and
Algorithm~\ref{alg:scc}. The results are shown in the right part of Table~\ref{tab:evaluation}.
Moreover, we also consider the network where the value of housekeeping is not fixed and show the
result in the row apoptosis*. When the housekeeping node is not fixed, the state-space of the
network is doubled. The results clearly indicate that our proposed decomposition method provides
better speedups with respect to Algorithm~\ref{alg:detect} for larger models.

\begin{table}[!t]
\centering
\begin{tabular}{lcccc|lcccc}
\toprule
\multicolumn{1}{c}{\multirow{2}{*}{Networks}} & \multirow{2}{*}{\begin{tabular}[c]{@{}c@{}}\# \\ attractors\end{tabular}} & \multicolumn{2}{c}{Time(s)}                              & \multirow{2}{*}{~Speedup}  & \multicolumn{1}{c}{\multirow{2}{*}{Networks}} & \multirow{2}{*}{\begin{tabular}[c]{@{}c@{}}\# \\ attractors\end{tabular}} & \multicolumn{2}{c}{Time(s)}                                & \multirow{2}{*}{~Speedup}   \\ \cline{3-4} \cline{8-9}
\multicolumn{1}{c}{}                         &                                                                          & Alg. 1                      & \multicolumn{1}{c}{Alg. 2} &                           & \multicolumn{1}{c}{}                         &                                                                          & Alg. 1                        & \multicolumn{1}{c}{Alg. 2} &                            \\ \hline
MAPK\_r3                                       & \multicolumn{1}{c}{20}                                                  & \multicolumn{1}{r}{6.070}  & 2.614 & \multicolumn{1}{r|}{\bf 2.32} & apoptosis                                     & \multicolumn{1}{c}{1024}                                                & \multicolumn{1}{c}{1633.970 } &   103.856                   & \multicolumn{1}{r}{\bf 15.73} \\
MAPK\_r4                                       & \multicolumn{1}{c}{24}                                                  & \multicolumn{1}{r}{11.674} & 1.949                       & \multicolumn{1}{r|}{\bf 5.99} & apoptosis*                                    & \multicolumn{1}{c}{2048}                                                & \multicolumn{1}{c}{8564.680} & 218.230                      & \multicolumn{1}{r}{\bf 39.25 } \\
 \bottomrule
\end{tabular}
\caption{Evaluation results on two real-life biological systems.}
\label{tab:evaluation}
\vspace{-6mm}
\end{table}

%===============================================
\section{Discussions and Future Work}
\label{sec:conc}
%===============================================
We have presented an~SCC-based decomposition method for detecting attractors in large asynchronous
BNs, which often arise and are important in the holistic study of biological systems. This problem
is very challenging as the state space of such networks is exponential in the number of nodes in
the networks and therefore huge. Meanwhile, asynchrony greatly increases the difficulty of
attractor detection as the density of the transition graph is inflated dramatically and the
structure of attractors may be complex. Our method performs SCC-based decomposition of the network
structure of a~give BN to manage the cyclic dependencies among network nodes, computes the
attractors of each SCC, and finally recovers the attractors of the original BN by merging the
detected (partial) attractors. To the best of our knowledge, our method is the first scalable one
able to deal with large biological systems modelled as asynchronous BNs, thanks to its divide and
conquer strategy. We have prototyped our method and performed experiments with two real biological
networks. The obtained results are very promising.

We have observed that the network structure of BNs can vary quite a~lot, which potentially has
impact on the performance of our proposed method. In principle, our method works well on large
networks which contain several relatively small SCCs. Each of the two mutants of the MAPK network,
however, contains one large SCC with 36 nodes and 17 SCCs each with one node only. Moreover, the
large SCC is in the middle of the SCC network structure (see Figure~\ref{fig:mapk_scc_structure}
in Appendix~\ref{app:mapk}). This network structure in fact does not fit well with our method.
This explains why the speedups achieved for this network are less than 10. Both the MAPK network
and the apoptosis network contain many small SCCs with only one node (see
Figure~\ref{fig:mapk_scc_structure} and Figure~\ref{fig:apoptosis_scc_structure}). One way to
improve our method is to merge these small SCCs into larger blocks so that there will be fewer
iterations in the main loop of Algorithm~\ref{alg:detect}. Moreover, the single-node SCCs which
do not have child SCCs are in fact leaves and they can be removed to reduce the network size. When
the attractors in the reduced network are detected, we can then recover the attractors in the
whole network.\footnote{This is in general related to network reduction techniques (e.g.,
see~\cite{SAA10}) which aim to simplify the networks prior to dynamic analysis.} Such
optimisations will be part of our future work. We will also apply our method to other realistic
large biological networks and we will develop optimisations fitted towards different SCC network 
structures.

\noindent{\bf Acknowledgement.}
Qixia Yuan is supported by the National Research Fund, Luxembourg (grant 7814267).

%===============================================
%----------The Bibliography--------------------
%===============================================
\bibliographystyle{splncs}
\bibliography{attractor}

\appendix
\newpage
%===============================================
\section{Additional definitions and lemmas used for proofs in Section~\ref{app:proofs}}
\label{app:defi}
%===============================================
\begin{definition}[Path, Hyper-path]
\label{def:path}
Given a~BN $G$ of $n$ nodes and its state space $X=\{0,1\}^n$,
%$X=\{\boldsymbol{x}_1,\boldsymbol{x}_2,\ldots,\boldsymbol{x}_{2^n}\}$,
a~\emph{path} of length $k~(k\! \geqslant\! 2)$ in $G$ is a~serial $\boldsymbol{x}_1 \rightarrow
\boldsymbol{x}_2\rightarrow\cdots \rightarrow \boldsymbol{x}_k$ of states in $X$ such that there
exists a~transition between any consecutive two states $\boldsymbol{x}_i$ and
$\boldsymbol{x}_{i+1}$, where $i \in [1,k-1]$.
A~\emph{hyper-path} of length $k~(k \geqslant 2)$ in $G$ is a~serial $\boldsymbol{x}_1
\dashrightarrow \boldsymbol{x}_2\dashrightarrow\cdots \dashrightarrow \boldsymbol{x}_k$ of states
in $X$ such that at least one of the two conditions is satisfied:
1)~there is a~transition from $\boldsymbol{x}_i$ to $\boldsymbol{x}_{i+1}$,
2)~$\boldsymbol{x}_i=\boldsymbol{x}_{i+1}$,
where $i \in [1,k-1]$.
\end{definition}

%Note that in the case of selfloops, any two consecutive states satisfy both conditions of
%the hyper-path. For definiteness, we treat this case as they satisfy the first condition.

The concepts of a~path and a~hyper-path in a~BN can be naturally extended to elementary blocks.
Notice that for any two consecutive states $\boldsymbol{x}_i$, $\boldsymbol{x}_{i+1}$ in a~path
$\boldsymbol{x}_1 \rightarrow \boldsymbol{x}_2\rightarrow\cdots \rightarrow \boldsymbol{x}_k$ in
a~BN, $k \geqslant 2$ and $i \in [1,k-1]$, if the transition between these two states is due to
the updating of a~node in an elementary block $B$, then there is a~transition from $\pi_B(\boldsymbol{x}_i)$ to
$\pi_B(\boldsymbol{x}_{i+1})$; otherwise, $\pi_B(\boldsymbol{x}_i)=\pi_B(\boldsymbol{x}_{i+1})$.
Therefore, the projection of all the states in the path $\boldsymbol{x}_1 \rightarrow
\boldsymbol{x}_2\rightarrow\cdots \rightarrow \boldsymbol{x}_k$ on block $B$ actually forms
a~hyper-path $\pi_B(\boldsymbol{x}_1) \dashrightarrow \pi_B(\boldsymbol{x}_2) \dashrightarrow
\cdots \dashrightarrow \pi_B(\boldsymbol{x}_k)$ in block $B$.
The following lemma follows immediately from the definitions of path and hyper-path.
\begin{lemma}
\label{lemma:hyperpath}
Let $\boldsymbol{x}_1 \dashrightarrow \boldsymbol{x}_2\dashrightarrow\cdots \dashrightarrow
\boldsymbol{x}_k$ be a~hyper-path in a~BN of length $k$. At least one of the two statements holds.
1)~There is a~path from $\boldsymbol{x}_1$ to $\boldsymbol{x}_k$ in the BN and this path contains
all the states in the hyper-path.
2)~$\boldsymbol{x}_1=\boldsymbol{x}_2 =\cdots =\boldsymbol{x}_k$.
\end{lemma}

\begin{lemma}
\label{lemma:atrractor_crossability}
Let ${B_j}$ be a~single, elementary parent block of a~non-elementary block $B_i$ in a~BN $G$. Let
$A^{B_j}$ be an~attractor of $B_j$ and let $A^{B_i}$ be an~attractor in the realisation of $B_i$
with respect to $A^{B_j}$. Then $A^{B_i} ~\mathcal{C}~ A^{B_j}$.
\end{lemma}
%===============================================
\section{Proofs}
\label{app:proofs}
%===============================================
%

\noindent
1) {\bf Proof of Lemma~\ref{lemma:dec-attractor}.}
\begin{proof}
Let $\mathcal{A}=\{A_1,A_2,\ldots,A_m\}$ be the
set of attractors of $G$ and $\mathcal{A}^{B}=\{A_1^{B},A_2^{B},\ldots,\allowbreak A_{m'}^{B}\}$
be the set of attractors of $B$. 
Since $B$ preserves the attractors of $G$, 
for any $k \in [1,m]$,
there exists a $k' \in [1,m']$ such that $\pi_B(A_k) \subseteq A_{k'}^{B}$.
Therefore, $\pi_B(\Phi)=\cup_{i=1}^{m} \pi_B(A_i) \subseteq \cup_{i=1}^{m'}A_i^{B}=\Phi^{B}$.
By Definition~\ref{def:mirror}, we have that $\Phi \subseteq \mathcal{M}_G(\pi_B(\Phi))$.
Hence, $\Phi \subseteq \mathcal{M}_G(\Phi^B)$.
\qed
\end{proof}

\medskip
\noindent
2) {\bf Proof of Theorem~\ref{theo:leavepres}.}

\begin{proof}
Let $\mathcal{A}=\{A_1,A_2,\ldots,A_m\}$ be the set of attractors of $G$.
%and $A^{B}=\{A_1^{B},A_2^{B},\cdots, A_{m^{B}}^{B}\}$ be attractors of $B$.
For any $i\in [1,m]$, let $L=\boldsymbol{x}_1\rightarrow \boldsymbol{x}_2 \rightarrow \cdots
\rightarrow \boldsymbol{x}_k$ be a~path containing all the states in $A_i$ and let
$\boldsymbol{x}_1=\boldsymbol{x}_k$. According to Definition~\ref{def:path},
$\pi_{B}(\boldsymbol{x}_1)\dashrightarrow\pi_{B}(\boldsymbol{x}_2)\dashrightarrow\cdots\dashrightarrow
\pi_{B}(\boldsymbol{x}_k)$ is a~hyper-path in $B$. We denote this hyper-path as $L^B$.
Therefore, one of the following two conditions must hold:
1)~there exists a~path $L'$ from $\pi_{B}(\boldsymbol{x}_1)$ to $\pi_{B}(\boldsymbol{x}_k)$ in B;
2)~$\pi_{B}(\boldsymbol{x}_1)=\pi_{B}(\boldsymbol{x}_2)=\cdots=\pi_{B}(\boldsymbol{x}_k)$.
Given that the choice of the attractor $A_i$ is arbitrary, the claim holds if we can prove that
states in the hyper-path $L^B$ form an~attractor of $B$ under both conditions. We will prove
them one by one.

\noindent\emph{Condition 1:} Given the arbitrary choice of the path, when the first condition
holds, the states in this path can reach each other. Now we only need to prove that the states in
this path cannot reach any other state that is not in this path. We prove by contradiction.
Assume a~state $\pi_B(\boldsymbol{x}_i)$ in path $L'$ can reach state $\pi_B(\boldsymbol{x}_i')$
by applying the Boolean function of some node $v_p$ and $\pi_B(\boldsymbol{x}_i')$ is not in $L'$.
Hence there is a~transition from $\boldsymbol{x}_i$ to $\boldsymbol{x}_i'$ in $G$. Since $L$
contains all the states in $A_i$ and $A_i$ is an~attractor, necessarily $\boldsymbol{x}_i'$ is
contained by $L$. Therefore, $\pi_B(\boldsymbol{x}_i')$ is one of the states in the hyper-path
$L^B$. According to Lemma~\ref{lemma:hyperpath}, all states in $L^B$ are contained by $L'$,
in particular $\pi_B(\boldsymbol{x}_i')$. This is contradictory to the assumption. It follows
that states of $L^B$ form an~attractor of $B$.

\noindent\emph{Condition 2:} This condition holds only when all transitions in path $L$ are
performed by applying Boolean functions of nodes that are not in block $B$. For any $j \in
[1,k-1]$, let $\boldsymbol{x}_{j'}$ be any state reachable from $\boldsymbol{x}_j$ by one
transition. We have $\boldsymbol{x}_{j'} \in A_i$ and therefore $L$ contains
$\boldsymbol{x}_{j'}$. Hence $L^B$ contains  $\pi_{B}(\boldsymbol{x}_{j'})$ and
$\pi_{B}(\boldsymbol{x}_{j'})=\pi_{B}(\boldsymbol{x}_1)=\pi_{B}(\boldsymbol{x}_2)=\cdots=
\pi_{B}(\boldsymbol{x}_k)$. Given the choice of $\boldsymbol{x}_j$ and $\boldsymbol{x}_{j'}$ is
arbitrary, $\pi_B(A_1)=\{\pi_B(\boldsymbol{x}_1)\}$, which is a~singleton attractor in $B$.
\qed
\end{proof}

\medskip
\noindent
3) {\bf Proof of Lemma~\ref{lemma:atrractor_crossability}.}
\begin{proof}
By the definition of realisation we have that for any state $\boldsymbol{x}^{B_i} \in A^{B_i}$,
there exists a~state $\boldsymbol{x}^{B_j} \in A^{B_j}$ such that $\boldsymbol{x}^{B_j}
~\mathcal{C}~ \boldsymbol{x}^{B_i}$.

Let us denote the set of control nodes of $B_i$ with $Z$, the set of the remaining nodes in $B_i$
with $V$, and use $\boldsymbol{zv}$ to represent a~state of block $B_i$ where $\boldsymbol{z}$ are
the values for the nodes in $Z$ and $\boldsymbol{v}$ are the values for the nodes in $V$. Let
$L^{B_j}$ be a~closed path, i.e., the first and the last state are the same, in the transition
system of $B_j$ which contains all the states in $A^{B_j}$. Let $\boldsymbol{x}^{B_i}$ be any
state in $A^{B_i}$. Due to the asynchronous updating scheme and the fact that the nodes in $Z$ are
independent of the nodes in $V$, one obtains that $\boldsymbol{z}\pi_V(\boldsymbol{x}^{B_i}) \in
A^{B_i}$ for any $\boldsymbol{z} \in \pi_Z(A^{B_j}) = \pi_Z(L^{B_j})$. For this it is enough to
observe that any of these states can be reached from $\boldsymbol{x}^{B_i}$ by following the
corresponding sequence of transitions in the hyper-path obtained by projecting $L^{B_j}$ on $Z$.
In consequence, for any state $\boldsymbol{x}^{B_j} \in A^{B_j}$ we have that
$\boldsymbol{x}^{B_j} ~\mathcal{C}~ \pi_Z(\boldsymbol{x}^{B_j})\pi_V(\boldsymbol{x}^{B_i})$ and
$\pi_Z(\boldsymbol{x}^{B_j})\pi_V(\boldsymbol{x}^{B_i}) \in A^{B_i}$. Hence,
$A^{B_i} ~\mathcal{C}~ A^{B_j}$.
\qed
\end{proof}

\medskip
\noindent
4) {\bf Proof of Theorem~\ref{theorem:recover}.}
\begin{proof}
We first prove that $\mathcal{A}^{B_i}~\mathcal{C}~\mathcal{A}^{B_j}$. If $B_i$ and $B_j$ are two
elementary blocks, they do not share common nodes. Then it holds by definition that
$\mathcal{A}^{B_i}~\mathcal{C}~\mathcal{A}^{B_j}$. Now, let $B_i$ be the only elementary parent
block of $B_j$. By definition, the attractors of $B_j$ is the set of the attractors of all
realisations of $B_j$. Due to this definition, for any attractor ${A}^{B_i} \in
\mathcal{A}^{B_i}$, one can always find an~attractor ${A}^{B_j} \in \mathcal{A}^{B_j}$ such that
${A}^{B_i}~\mathcal{C}~{A}^{B_j}$. For this it is enough to consider the realisation of $B_j$
with respect to ${A}^{B_i}$ and to take as ${A}^{B_j}$ one of the attractors of this realisation.
By Lemma~\ref{lemma:atrractor_crossability} we have that ${A}^{B_i}~\mathcal{C}~{A}^{B_j}$.
Further, again by Lemma~\ref{lemma:atrractor_crossability}, for any attractor ${A}^{B_j} \in
\mathcal{A}^{B_j}$, there is an~attractor ${A}^{B_i} \in \mathcal{A}^{B_i}$ such that ${A}^{B_i}~
\mathcal{C}~ {A}^{B_j}$, i.e., the one that gives rise to the realisation of which ${A}^{B_j}$ is
an~attractor in $B_j$. Therefore, $\mathcal{A}^{B_i}~\mathcal{C}~\mathcal{A}^{B_j}$.

We now prove that $\Pi(\mathcal{A}^{B_i},\mathcal{A}^{B_j})$ is the set of attractors of
$B_{i,j}$. This is equivalent to showing the following two statements:
1)~for any $A \in \Pi(\mathcal{A}^{B_i},\mathcal{A}^{B_j})$, $A$ is an~attractor of $B_{i,j}$;
2)~any attractor of $B_{i,j}$ is contained in $\Pi(\mathcal{A}^{B_i},\mathcal{A}^{B_j})$.
We prove them one by one.

\smallskip
\noindent\emph{Statement~1:} Let $A$ be any set of states in
$\Pi(\mathcal{A}^{B_i},\mathcal{A}^{B_j})$. Then there exist $A^{B_i} \in \mathcal{A}^{B_i}$ and
$A^{B_j} \in \mathcal{A}^{B_j}$ such that $A=\Pi(A^{B_i},A^{B_j})$ and
$A^{B_i}~\mathcal{C}~A^{B_j}$. We first prove that $\boldsymbol{x} =
\Pi(\pi_{B_i}(\boldsymbol{x}),\pi_{B_j}(\boldsymbol{x}))$, where $\pi_{B_i}(\boldsymbol{x}) \in
A^{B_i}$ and $\pi_{B_j}(\boldsymbol{x}) \in A^{B_j}$, cannot reach any state that is not in $A$ by
contradiction. Assume that $\boldsymbol{x}$ can reach a~state $\boldsymbol{y}$ by one
transition and $\boldsymbol{y} \notin A$. Due to asynchronous updating mode, the transition from
$\boldsymbol{x}$ to $\boldsymbol{y}$ is caused by updating one node. There are three
possibilities: 1)~the updated node is in $B_i$ and it is not a~control node of $B_j$; 2)~the
updated node is in $B_j$ and it is not a~control node; 3)~the updated node is a~control node of
$B_j$. In the first case, there is a~transition from $\pi_{B_i}(\boldsymbol{x})$ to
$\pi_{B_i}(\boldsymbol{y})$ in the elementary block $B_i$ and since $\pi_{B_i}(\boldsymbol{x})$
belongs to attractor $A^{B_i}$, it follows that $\pi_{B_i}(\boldsymbol{y}) \in A^{B_i}$.
In addition, we have $\pi_{B_j}(\boldsymbol{y})=\pi_{B_j}(\boldsymbol{x})$.
Then $\boldsymbol{y} = \Pi(\pi_{B_i}(\boldsymbol{y}),\pi_{B_j}(\boldsymbol{x}))$ and $\boldsymbol{y}
\in A$. Similarly in the second case, there is a~transition from $\pi_{B_j}(\boldsymbol{x})$ to
$\pi_{B_j}(\boldsymbol{y})$ within the attractor system $A^{B_j}$, so
$\pi_{B_j}(\boldsymbol{y}) \in A^{B_j}$ and $\boldsymbol{y} =
\Pi(\pi_{B_i}(\boldsymbol{x}),\pi_{B_j}(\boldsymbol{y})) \in A$. In the third case, there is
a~transition from $\pi_{B_i}(\boldsymbol{x})$ to $\pi_{B_i}(\boldsymbol{y})$ in the elementary
block $B_i$ and, as in the first case, we have that $\pi_{B_i}(\boldsymbol{y}) \in A^{B_i}$.
Since $A^{B_i}~\mathcal{C}~A^{B_j}$, there exists $\boldsymbol{s} \in A^{B_j}$ such
that $\pi_{B_i}(\boldsymbol{y})~\mathcal{C}~\boldsymbol{s}$. Let us denote the set of control
nodes of $B_j$ with $Z$, the set of the remaining nodes in $B_j$ with $V$, and use
$\boldsymbol{zv}$ to represent a~state of block $B_j$ where $\boldsymbol{z}$ are the values for
the nodes in $Z$ and $\boldsymbol{v}$ are the values for the nodes in $V$. Now, $\boldsymbol{s} =
\pi_{Z}(\boldsymbol{s})\pi_{V}(\boldsymbol{s})$ and there is a~path from
$\pi_{B_j}(\boldsymbol{x})$ to $\boldsymbol{s}$ in the attractor system $A^{B_j}$ as both states
belong to $A^{B_j}$. Since at each step of this path the value of only a~single node
is updated and the the control nodes in $Z$ are updated independently of the nodes in $V$, it
follows that starting from $\pi_{B_j}(\boldsymbol{x})$ and by following only the updates related
to the control nodes in $Z$ in the path from $\pi_{B_j}(\boldsymbol{x})$ to $\boldsymbol{s}$,
there is a~path in the attractor system $A^{B_j}$ from $\pi_{B_j}(\boldsymbol{x})$ to
$\pi_{Z}(\boldsymbol{s})\pi_{V}(\boldsymbol{x}) = \pi_{B_j}(\boldsymbol{y})$. Hence,
$\pi_{B_j}(\boldsymbol{y}) \in A^{B_j}$ and we have that $\boldsymbol{y} =
\Pi(\pi_{B_i}(\boldsymbol{y}),\pi_{B_j}(\boldsymbol{y})) \in A$. In all three cases we reach
a~contradiction.

We now show that for any two states $\boldsymbol{a},\boldsymbol{x} \in A=\Pi(A^{B_i},A^{B_j})$,
$\boldsymbol{x}$ is reachable from $\boldsymbol{a}$ only via states in $A$. We have
$\pi_{B_i}(\boldsymbol{a})$, $\pi_{B_i}(\boldsymbol{x}) \in A^{B_i}$ and there is a~path $L^{B_i}$
from $\pi_{B_i}(\boldsymbol{a})$ to $\pi_{B_i}(\boldsymbol{x})$ in $A^{B_i}$. Similarly, there is
a~path $L^{B_j}$ from $\pi_{B_j}(\boldsymbol{a})$ to $\pi_{B_j}(\boldsymbol{x})$ in the attractor
system of $A^{B_j}$. Following the same updating rules as in the path $L^{B_i}$, there is a~path
$L_1^{B_{i,j}}$ in $B_{i,j}$ from state $\boldsymbol{a}$ to state $\boldsymbol{y}$ such that
$\pi_{B_i}(\boldsymbol{y})=\pi_{B_i}(\boldsymbol{x})$ and the non-control nodes of $B_j$ in
$\boldsymbol{y}$ have the same values as in $\boldsymbol{a}$. The claim holds if we can prove that
there is a~path $\overline{L^{B_j}}$ in the attractor system of $A^{B_j}$ from state
$\pi_{B_j}(\boldsymbol{y})$ to $\pi_{B_j}(\boldsymbol{x})$ since following the same updating rules
as in the path $\overline{L^{B_j}}$, there is a~path $L_2^{B_{i,j}}$ in $B_{i,j}$ from
$\boldsymbol{y}$ to $\boldsymbol{x}$ and hence $\boldsymbol{x}$ is reachable from
$\boldsymbol{a}$. We prove this in the following two cases. The first case is when $B_i$ and $B_j$
are both elementary blocks. In this case, the merged block $B_{i,j}$ is in fact a~$BN$ and we have
$\pi_{B_j}(\boldsymbol{a})=\pi_{B_j}(\boldsymbol{y})$. Therefore, the path $L^{B_j}$ is in fact
$\overline{L^{B_j}}$. We now consider the second case where $B_i$ is a~parent of $B_j$. Using the
notation introduced above, we show that the state $\pi_{B_j}(\boldsymbol{y}) =
\pi_Z(\boldsymbol{y})\pi_V(\boldsymbol{a}) \in A^{B_j}$. This follows from applying the
corresponding argumentation for node update possibilities one or three presented above at each
step of the path $L^{B_i}$. Now, since both $\pi_{B_j}(\boldsymbol{y})$ and
$\pi_{B_j}(\boldsymbol{x})$ belong to $A^{B_j}$, there is a~path from $\pi_{B_j}(\boldsymbol{y})$
to $\pi_{B_j}(\boldsymbol{x})$ in the attractor system of $A^{B_j}$. This path is exactly the
searched path $\overline{L^{B_j}}$. Given the choice of $\boldsymbol{a}$ and $\boldsymbol{x}$ is
arbitrary, we can claim that any two states in $A$ are reachable from each other. Moreover, since
a~state in $A$ cannot reach any state outside $A$ as shown above, the two states in $A$ are
reachable from each other via states only in $A$. Hence, Statement~1 follows.

\smallskip
\noindent\emph{Statement~2:} We prove that $\Pi(\mathcal{A}^{B_i},\mathcal{A}^{B_j})$ contains all
the attractors of $B_{i,j}$. Let $A^{B_{i,j}}$ be an~attractor in $B_{i,j}$. Since the nodes in
$B_i$ are independent of the nodes in $B_j$, clearly $\pi_{B_i}(A^{B_{i,j}})$ is an~attractor in
$B_i$. Therefore, $\pi_{B_i}(A^{B_{i,j}}) \in \mathcal{A}^{B_i}$.

Let us consider the realisation of block $B_j$ with respect to $\pi_{B_i}(A^{B_{i,j}})$. We
proceed to show that $\pi_{B_j}(A^{B_{i,j}})$ is an~attractor of this realisation. Let us assume
that there exists $\boldsymbol{x} \in \pi_{B_j}(A^{B_{i,j}})$ such that it can reach a~state
$\boldsymbol{y} \notin \pi_{B_j}(A^{B_{i,j}})$ by one transition in the realisation. Let
$\boldsymbol{\tilde{x}} \in A^{B_{i,j}}$ be the corresponding state of $\boldsymbol{x}$ in
$A^{B_{i,j}}$, i.e. $\pi_{B_j}(\boldsymbol{\tilde{x}}) = \boldsymbol{x}$. It follows that there
exists a~state $\boldsymbol{\tilde{y}}$ of $B_{i,j}$ reachable from $\boldsymbol{\tilde{x}}$ by
one transition such that $\pi_{B_j}(\boldsymbol{\tilde{y}}) = \boldsymbol{y}$. In consequence,
$\boldsymbol{\tilde{y}} \notin A^{B_{i,j}}$ and $A^{B_{i,j}}$ cannot be an~attractor. This
contradicts the original assumption.

Now we show that there is a~path between any two states $\boldsymbol{x}$ and $\boldsymbol{y}$ of
$\pi_{B_j}(A^{B_{i,j}})$ in the realisation only via states in $\pi_{B_j}(A^{B_{i,j}})$. The
existence of such path follows in a~straightforward way from the fact that there exist two
corresponding states $\boldsymbol{\tilde{x}}$, $\boldsymbol{\tilde{y}}$ in $A^{B_{i,j}}$ such that
$\pi_{B_j}(\boldsymbol{\tilde{x}}) = \boldsymbol{x}$ and $\pi_{B_j}(\boldsymbol{\tilde{y}}) =
\boldsymbol{y}$. In consequence, there is a~path from one to the other as both are in
the attractor $A^{B_{i,j}}$. Projection of this path on $B_j$ forms a~hyper-path in the
realisation. By Lemma~\ref{lemma:hyperpath}, $\boldsymbol{y}$ is reachable from $\boldsymbol{x}$
in the realisation and only via states in $\pi_{B_j}(A^{B_{i,j}})$ as shown above. Hence,
$\pi_{B_j}(A^{B_{i,j}})$ is an~attractor of the considered realisation, i.e.
$\pi_{B_j}(A^{B_{i,j}}) \in \mathcal{A}^{B_j}$.

Finally, it is straightforward to verify that
$\pi_{B_i}(A^{B_{i,j}}) ~\mathcal{C}~ \pi_{B_j}(A^{B_{i,j}})$. Therefore, $A \in
\Pi(\mathcal{A}^{B_i},\mathcal{A}^{B_j})$, which concludes the proof of Statement~2 and the
theorem.
\qed
\end{proof}

\medskip
\noindent
5) {\bf Proof of Corollary~\ref{corollary:order}.}
\begin{proof}
According to Theorem~\ref{theorem:recover}, $\Pi(\mathcal{A}^{B_i},\mathcal{A}^{B_j}$) is the set
of attractors of $B_{i,j}$ and $\Pi(\mathcal{A}^{B_i},\mathcal{A}^{B_k})$ is the set of attractors
of $B_{i,k}$. Merging $B_i$ with $B_j$ results in an~elementary block $B_{i,j}$, and merging $B_i$
with $B_k$ results in an~elementary block $B_{i,k}$. Applying Theorem~\ref{theorem:recover} again,
we get $\Pi(\Pi(\mathcal{A}^{B_i},\mathcal{A}^{B_j}),\mathcal{A}^{B_k})$ is the set of attractors
of $B_{i,j,k}$ and $\Pi(\Pi(\mathcal{A}^{B_i},\mathcal{A}^{B_k}),\mathcal{A}^{B_j})$ is the set of
attractors of $B_{i,k,j}$. Since $B_{i,j,k}$ and $B_{i,k,j}$ are actually the same block,
$\Pi(\Pi(\mathcal{A}^{B_i},\mathcal{A}^{B_j}),\mathcal{A}^{B_k})=
\Pi(\Pi(\mathcal{A}^{B_i},\mathcal{A}^{B_k}),\mathcal{A}^{B_j})$.
\qed
\end{proof}

%We now prove that Algorithm~\ref{alg:scc} can identify all the attractors of a given BN $G$.
\medskip
\noindent
6) {\bf Proof of Theorem~\ref{theo:scc-identify}.}
\begin{proof}
Algorithm~\ref{alg:scc} divides a BN into SCC blocks and detects attractors of each block.
Line~\ref{line:detectblockstart} to~\ref{line:detectblockend} describe the process for detecting attractors of a block.
The algorithm distinguishes between two different types of blocks.
The first type is an elementary block.
Since it is in fact a BN, the attractors of this type of block are directly detected via Algorithm~\ref{alg:detect}.
The second type is a non-elementary block.
The algorithm constructs the realisations of this type of block,
detects attractors of each realisation and merges them as the attractors of the block.
The algorithm takes special care of those blocks with more than one parent blocks.
It merges all the ancestor blocks of such a block as its parent block.
Since the ancestor blocks are in ascending operations based on their credits,
the cross operation in Line~\ref{line:api} will iteratively recover the attractors of the parent block according to Theorem~\ref{theorem:recover}.
Whenever the attractors of a block $B_i$ are detected,
it performs a cross operation between block $B_i$ and the elementary block $B_c$ formed by nodes in all previous blocks (Line~\ref{line:cross}).
According to Theorem~\ref{theorem:recover},
the cross operation will result in the attractors of the block formed by nodes in the two blocks.
Since Algorithm~\ref{alg:scc} iteratively performs this operation to all the blocks,
it will recover the attractors of the BN in the last iteration.
Note that how to order two blocks with the same credit does not affect the result of this algorithm,
as proved in Corollary~\ref{corollary:order}.
\qed
\end{proof}

%\noindent
%7) Proof for Theorem~\ref{theo:leafalgo}.
%\begin{proof}
%Algorithm~\ref{alg:leaf} only searches the states in the set $\mathcal{M}_G(\Phi^{B})$ for
%attractors of $G$. According to Lemma~\ref{lemma:dec-attractor} and Theorem~\ref{theo:leavepres},
%$\mathcal{M}_G(\Phi^{B})$ contains all the attractor states of $G$. Therefore, by searching
%$\mathcal{M}_G(\Phi^{B})$ Algorithm~\ref{alg:leaf} can identify all the attractor states of $G$.
%\qed
%\end{proof}
\newpage
%===============================================
\section{The Boolean Model of MAPK network}
\label{app:mapk}
%===============================================
\begin{figure}[!ht]
\centering
\includegraphics[width=.98\textwidth]{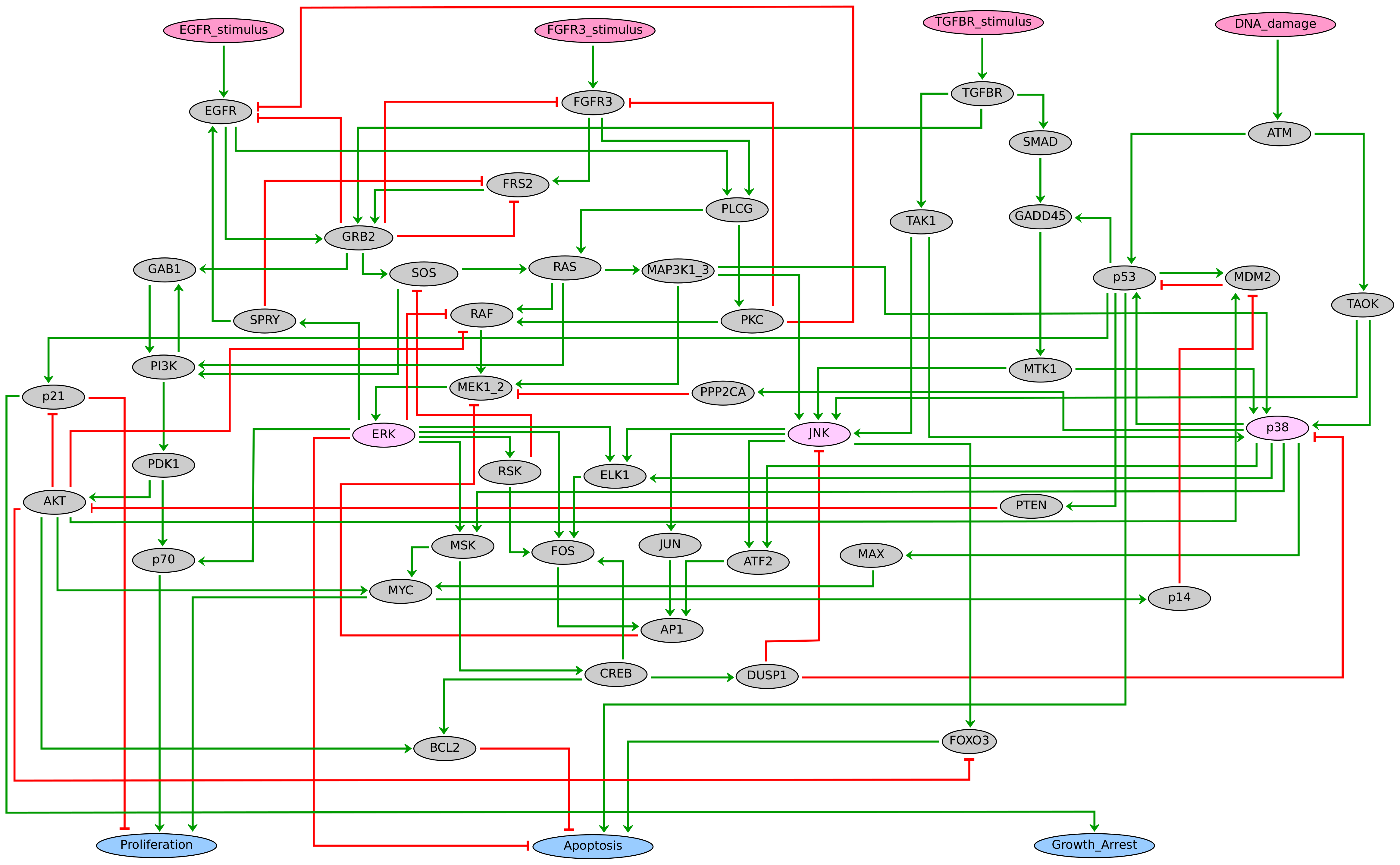}
\caption{Wiring of the MAPK logical model of~\cite{GCT13}. The diagram contains three types of
nodes: stimuli nodes (pink), signalling component nodes (gray) with highlighted MAPK protein
nodes (light pink), and cell fate nodes (blue). Green arrows and red blunt arrows represent
positive and negative regulations, respectively. For detailed information on the Boolean model of
the MAPK network containing all modelling assumptions and specification of the logical rules refer
to~\cite{GCT13} and the supplementary material thereof.}
\label{fig:mapk_structure}
\end{figure}

\newpage

\begin{figure}[!ht]
\centering
\includegraphics[width=0.55\textwidth]{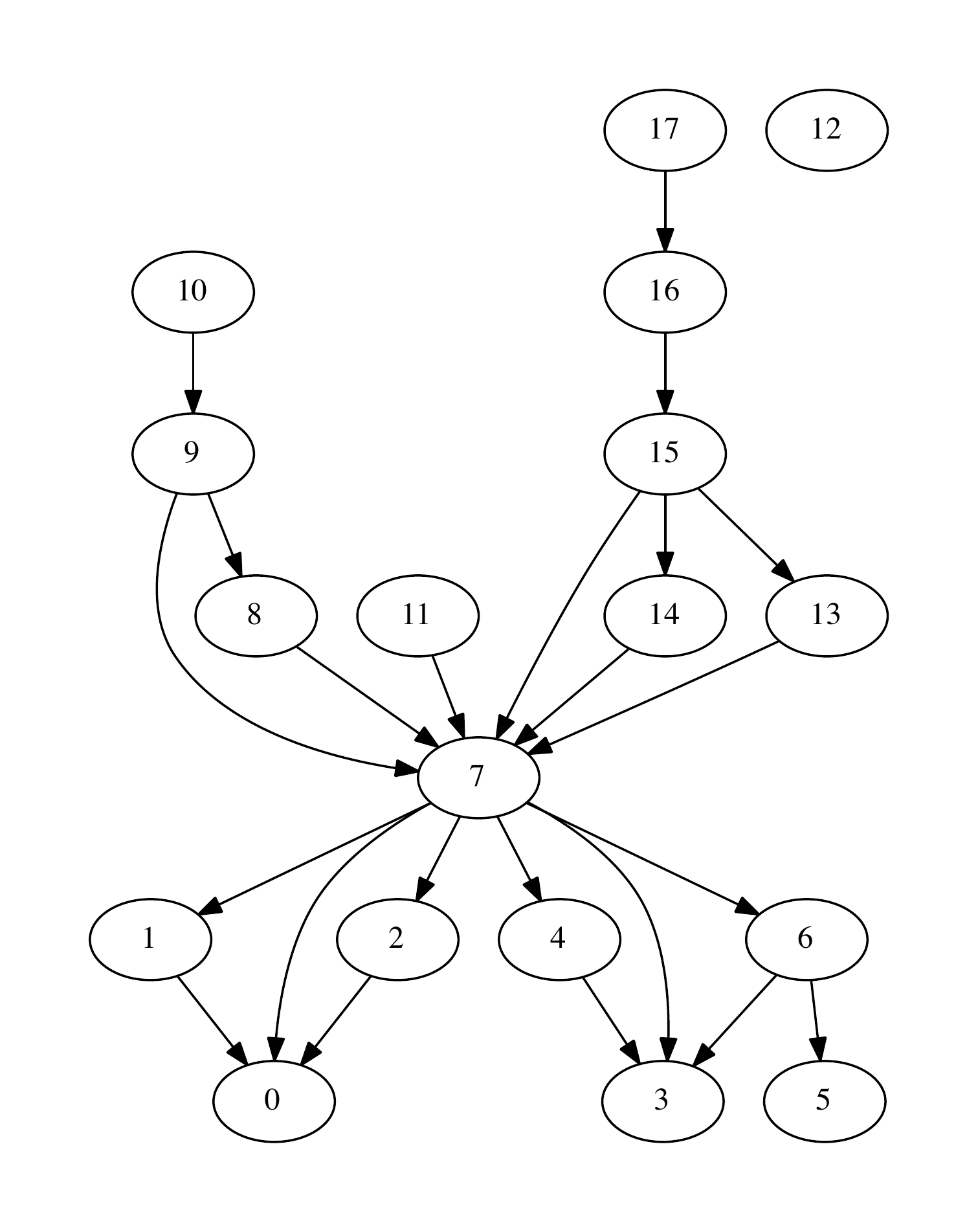}
\caption{The SCC structure of the MAPK network (mutant MAPK\_r3). Each node represents an~SCC.
Model components contained in each SCC are listed in Table~\ref{tab:sccnodes_mapk}. For each pair
of a~parent SCC and one of its child SCCs, a~directed edge is drawn pointing from the parent SCC
to the child SCC. Node 12 is not connected to any other node as EGFR is set to be always true and
hence the influence from EGFR\_stimulus (node 12) is cut. The SCC structure of mutant MAPK\_r4 is
virtually the same; the only difference is that model components contained in certain SCCs are
slightly different: EGFR is switched with FGFR3 and EGFR\_stimulus is switched with
FGFR3\_stimulus.}
\label{fig:mapk_scc_structure}
\end{figure}

% Please add the following required packages to your document preamble:
% \usepackage{multirow}
\begin{table}[!ht]
\centering
\begin{tabular}{|r|l|r|l|r|l|r|l|}
\hline
\multicolumn{1}{|c|}{scc \#} & \multicolumn{1}{c|}{nodes}              & \multicolumn{1}{c|}{scc \#}              & \multicolumn{1}{c|}{nodes}              & \multicolumn{1}{c|}{scc \#}              & \multicolumn{1}{c|}{nodes}              & \multicolumn{1}{c|}{scc \#}             & \multicolumn{1}{c|}{nodes}             \\ \hline
0                            & Apoptosis                               & 4                                        & p70                                     & 9                                        & ATM                                     & 13                                      &        FGFR3\_stimulus                \\ \hline
1                            & BCL2                                    & 5                                        & Growth\_Arrest                          & 10                                       & DNA\_damage                             & 14                                      &          SMAD                         \\ \hline

2 & FOXO3 & 6   & p21 & 11  & EGFR& 15 &TAK1    \\ \hline
3  & Proliferation  & 8  & TAOK  & 12  & EGFR\_stimulus & 16  & TGFBR    \\ \hline

17                           & TGFR\_stimulus                         &                                      & & &&  &  \\ \hline
\multirow{3}{*}{7}           & \multicolumn{7}{l|}{\multirow{3}{*}{\begin{tabular}[c]{@{}l@{}}AKT  AP1  ATF2  CREB  DUSP1   FGFR3  ELK1  ERK  FOS  FRS2  GAB1  \\GADD45  GRB2 JNK  JUN  MAP3K1\_3  MAX  MDM2  MEK1\_2  MSK  MTK1  MYC  \\PDK1   PI3K  PKC  PLCG  PPP2CA  PTEN  RAF  RAS  RSK  SOS  SPRY  p14  p38  p53\end{tabular}}} \\
                             & \multicolumn{7}{l|}{}                                                                                                                                                                                                                                                                                \\
                             & \multicolumn{7}{l|}{}                                                                                                                                                                                                                                                                                \\ \hline
\end{tabular}
\caption{Nodes of the MAPK pathway (mutant r3) in SCCs as shown in Figure~\ref{fig:mapk_scc_structure}.}
\label{tab:sccnodes_mapk}
\end{table}

\newpage
%===============================================
\section{The Boolean Model of Apoptosis}
\label{app:apoptosis}
%===============================================
\begin{figure}[!ht]
\centering
\includegraphics[width=.9\textwidth,angle=0]{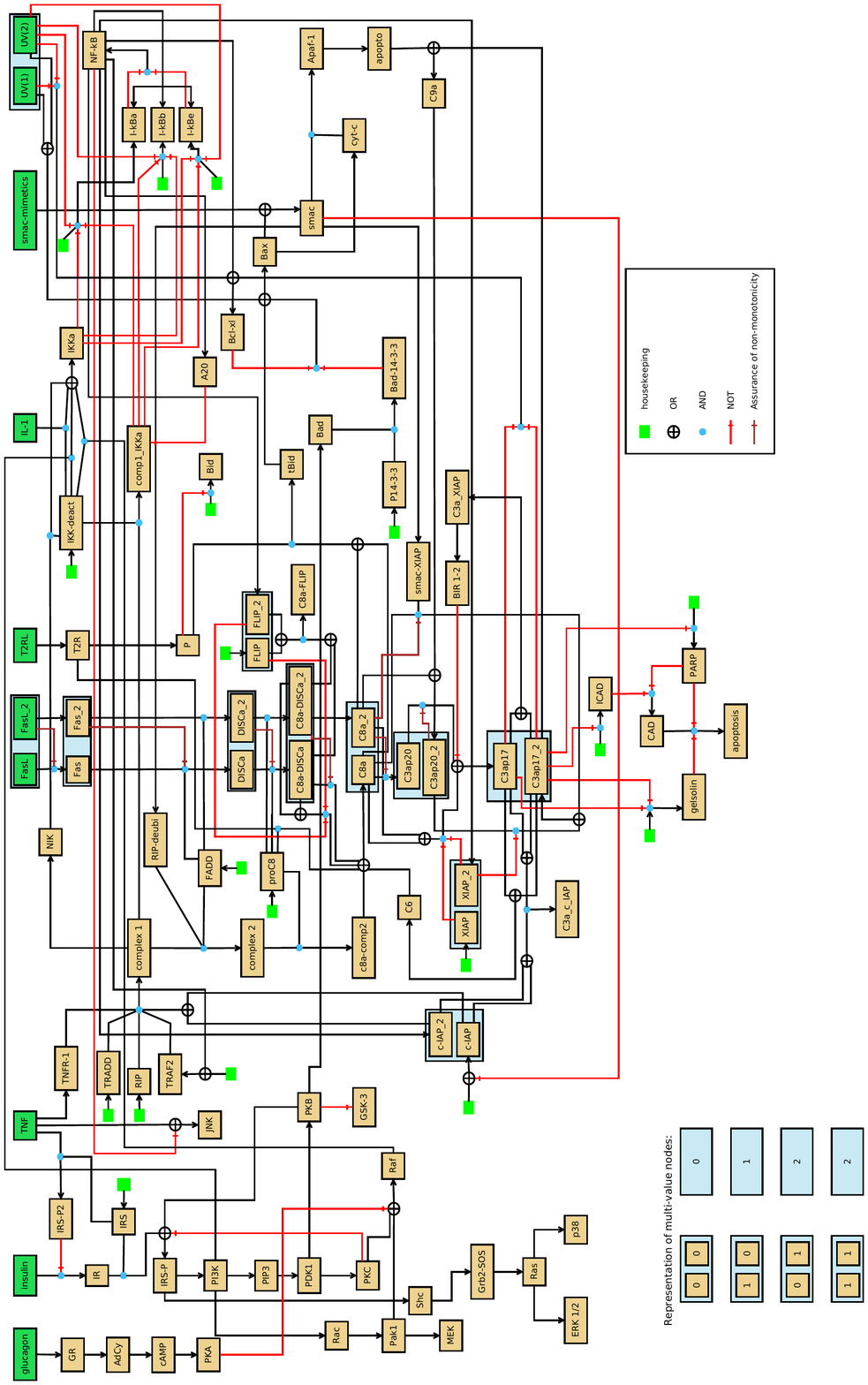}
\caption{The wiring of the multi-value logic model of apoptosis by Schlatter et
al.~\cite{SSVSSBEMS09} recast into a~binary Boolean network. For clarity of the diagram the nodes
I-kBa, I-kBb, and I-kBe have two positive inputs. The inputs are interpreted as connected via
$\oplus$ (logical OR).}
\label{fig:structure}
\end{figure}

\begin{figure}[!ht]
\centering
\includegraphics[width=1.3\textwidth,angle=90]{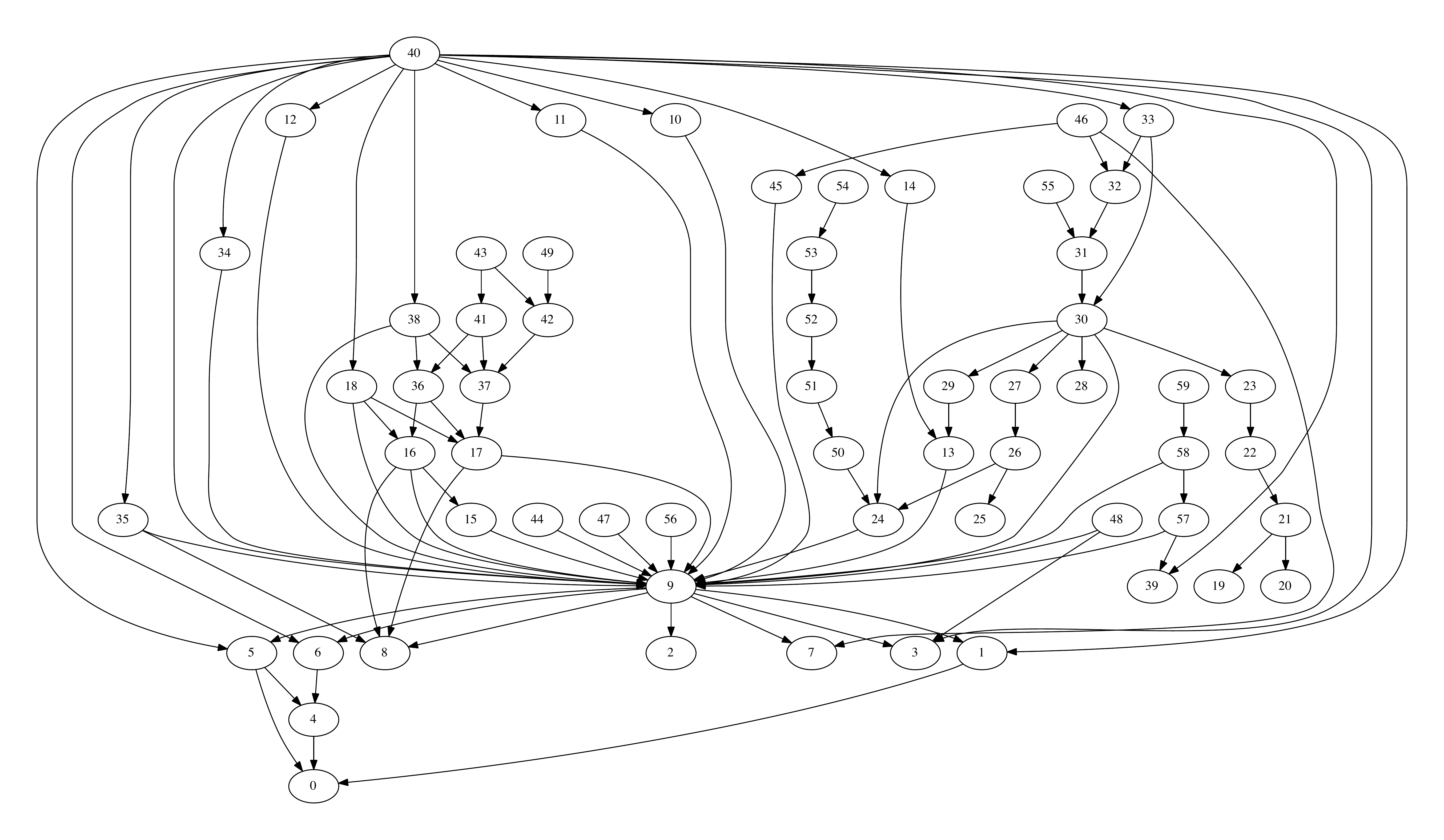}
\caption{The SCC structure of the apoptosis model. Each node represents an SCC in the apoptosis model.
The nodes contained in each SCC are listed in Table~\ref{tab:sccnodes_apoptosis}.
For each pair of a parent SCC and one of its child SCCs,
a directed edge is added pointing from the parent SCC to the child SCC.}
\label{fig:apoptosis_scc_structure}
\end{figure}

\begin{table}[]
\centering
\begin{tabular}{|r|l|r|l|r|l|r|l|}
\hline
\multicolumn{1}{|c|}{scc \#} & \multicolumn{1}{c|}{nodes} & \multicolumn{1}{c|}{scc \#} & \multicolumn{1}{c|}{nodes} & \multicolumn{1}{c|}{scc \#} & \multicolumn{1}{c|}{nodes} & \multicolumn{1}{c|}{scc \#} & \multicolumn{1}{c|}{nodes} \\ \hline
0                            & apoptosis                  & 16                          & C8a\_DISCa\_2              & 32                          & IRS\_P2                    & 47                          & UV                         \\ \hline
1                            & gelsolin                   & 17                          & C8a\_DISCa                 & 33                          & IRS                        & 48                          & UV\_2                      \\ \hline
2                            & C3a\_c\_IAP                & 18                          & proC8                      & 34                          & IKKdeact                   & 49                          & FASL                       \\ \hline
3                            & I\_kBb                     & 19                          & p38                        & 35                          & FLIP                       & 50                          & PKA                        \\ \hline
4                            & CAD                        & 20                          & ERK1o2                     & 36                          & DISCa\_2                   & 51                          & cAMP                       \\ \hline
5                            & PARP                       & 21                          & Ras                        & 37                          & DISCa                      & 52                          & AdCy                       \\ \hline
6                            & ICAD                       & 22                          & Grb2\_SOS                  & 38                          & FADD                       & 53                          & GR                         \\ \hline
7                            & JNK                        & 23                          & Shc                        & 39                          & Bid                        & 54                          & Glucagon                   \\ \hline
8                            & C8a\_FLIP                  & 24                          & Raf                        & 40                          & housekeeping               & 55                          & Insulin                    \\ \hline
10                           & XIAP                       & 25                          & MEK                        & 41                          & FAS\_2                     & 56                          & smac\_mimetics             \\ \hline
11                           & TRADD                      & 26                          & Pak1                       & 42                          & FAS                        & 57                          & P                          \\ \hline
12                           & RIP                        & 27                          & Rac                        & 43                          & FASL\_2                    & 58                          & T2R                        \\ \hline
13                           & Bad\_14\_3\_3              & 28                          & GSK\_3                     & 44                          & IL\_1                      & 59                          & T2RL                       \\ \hline
14                           & P14\_3\_3                  & 29                          & Bad                        & 45                          & TNFR\_1                    &                             &                            \\ \hline
15                           & C8a\_2                     & 31                          & IR                         & 46                          & TNF                        &                             &                            \\ \hline
\multicolumn{1}{|r|}{\multirow{4}{*}{9}} & \multicolumn{7}{l|}{\multirow{4}{*}{\begin{tabular}[c]{@{}l@{}}Apaf\_1 apopto A20 Bax Bcl\_xl BIR1\_2 c\_IAP c\_IAP\_2 complex1 \\ comp1\_IKKa cyt\_c C3ap20 C3ap20\_2 C3a\_XIAP C8a\_comp2 C9a \\ FLIP\_2 NIK RIP\_deubi smac smac\_XIAP tBid TRAF2 XIAP\_2 IKKa\\ I\_kBa I\_kBe complex2 NF\_kB C8a C3ap17 C3ap17\_2\end{tabular}}} \\
\multicolumn{1}{|r|}{}                   & \multicolumn{7}{l|}{}                                                                                                                                                                                                                                                                                                                 \\
\multicolumn{1}{|r|}{}                   & \multicolumn{7}{l|}{}                                                                                                                                                                                                                                                                                                                 \\
\multicolumn{1}{|r|}{}                   & \multicolumn{7}{l|}{}                                                                                                                                                                                                                                                                                                                 \\ \hline
\multicolumn{1}{|r|}{30}                 & \multicolumn{7}{l|}{IRS\_P PDK1 PKB PKC PIP3 PI3K C6}                                                                                                                                                                                                                                                                                 \\ \hline

\end{tabular}
\caption{Nodes of the apoptosis network in SCCs as shown in Figure~\ref{fig:apoptosis_scc_structure}.}
\label{tab:sccnodes_apoptosis}
\end{table}

\end{document}